\documentclass[aps,prb,twocolumn,longbibliography]{revtex4-2}

\usepackage{amsfonts}
\usepackage{amsmath}
\usepackage{graphicx}
\usepackage{dsfont}
\usepackage{diagbox}
\usepackage{array}
\usepackage{amssymb}
\usepackage{bbm}
\usepackage{float}
\usepackage{subfigure}
\usepackage{MnSymbol}
\usepackage{url}
\usepackage{times}
\usepackage{manfnt}
\usepackage{booktabs}
\usepackage{xcolor}
\definecolor{darkblue}{HTML}{004D6B}
\definecolor{darkred}{HTML}{8c1515}
\definecolor{darkgreen}{HTML}{006400}
\usepackage{hyperref}
\hypersetup{
    pdftitle={FractionalizedFloquetCode},
	colorlinks=true,
	urlcolor=darkred,
	citecolor=darkblue,
	linkcolor=darkred,
	breaklinks
}
\pagecolor{white}
\usepackage{ulem}
\usepackage{physics}
\usepackage{enumitem}
\usepackage{wasysym}

\begin{document}

\title{Qubit fractionalization and emergent Majorana liquid in the honeycomb Floquet code\\ induced by coherent errors and weak measurements}
\author{Guo-Yi Zhu}
\email{gzhu@uni-koeln.de}
\affiliation{Institute for Theoretical Physics, University of Cologne, Z\"ulpicher Straße 77, 50937 Cologne, Germany}

\author{Simon Trebst}
\affiliation{Institute for Theoretical Physics, University of Cologne, Z\"ulpicher Straße 77, 50937 Cologne, Germany}

\date{\today}

\begin{abstract}
From the perspective of quantum many-body physics, the Floquet code of Hastings and Haah can be thought of 
as a measurement-only version of the Kitaev honeycomb model where a periodic sequence of two-qubit XX, YY, and ZZ measurements dynamically stabilizes a toric code state with two logical qubits.
However, the most striking feature of the Kitaev model is its intrinsic fractionalization of quantum spins into
an emergent gauge field and itinerant Majorana fermions that form a Dirac liquid, which is absent in the  Floquet code. 
Here we demonstrate that by varying the measurement strength of the honeycomb Floquet code one can observe 
features akin to the fractionalization physics of the Kitaev model at finite temperature. 
Introducing coherent errors to weaken the measurements we observe three consecutive stages that reveal
qubit fractionalization (for weak measurements), the formation of a Majorana liquid (for intermediate measurement strength), 
and Majorana pairing together with gauge ordering (for strong measurements).
Our analysis is based on a mapping of the imperfect Floquet code to random Gaussian fermionic circuits (networks) that can be Monte Carlo sampled, exposing  two crossover peaks. 
With an eye on circuit implementations, our analysis demonstrates that the Floquet code, in contrast to the toric code, does not immediately 
break down to a trivial state under weak measurements, but instead gives way to a long-range entangled Majorana liquid state.
\end{abstract}

\maketitle


In the theory of quantum error correction, it has long been appreciated that measurements can not only extract
information from a quantum system, but they can also play a converse role in protecting quantum information.
This latter idea has been embodied in {\it stabilizer codes} \cite{gottesman1997stabilizer}, such as Kitaev's toric code \cite{Kitaev2003},
which allow to encode logical qubit(s) in a larger number of system qubits via a set of commuting measurement operations.
More recently, these concepts have been expanded by the introduction of 
{\it Floquet codes}~\cite{Haah21honeycomb,Hastings22honeycomb, Wootton22floquetcodeibm, Kesselring22colorcode, Nat23floquetwithoutparent, Vu23Floquetcode, Potter23Floquetcode, Roger23Floquetcode, Dua23floquetcodetwist}, 
a distinct class of error correcting codes 
that dynamically stabilize logical qubits via a time-periodic sequence of {\it non-commuting} measurement operations. 
The principal example of Hastings and Haah \cite{Haah21honeycomb}, which shows that a crucial benefit of such a dynamical approach
is that the required measurements can be reduced to {\it two-qubit} measurements, relies on three consecutive rounds of
XX, YY, and ZZ checks along the bonds of the honeycomb lattice.
As such this system bears a fundamental connection to the Kitaev honeycomb model \cite{Kitaev2006}, which has been widely studied
for its spin liquid ground states \cite{Hermanns2018}, arising from the fractionalization of quantum spins as a result of the frustration induced by the 
competition of non-commuting terms in the Hamiltonian.
The honeycomb Floquet code of Hastings and Haah connects to this physics by stabilizing a gapped toric code-like state
in each individual measurement round, akin to the 
strong bond anisotropy limit of the Kitaev model phase diagram. 
With the code's three-fold time and space periodicity a non-trivial dynamics ensues 
\cite{Hastings22honeycomb, Vu23Floquetcode, Potter23Floquetcode, Roger23Floquetcode}, 
which in the Hamiltonian language corresponds to jumping between the topological ground-state phases in the corners 
of its triangular phase diagram \cite{Schmidt10honeycomb}.
The honeycomb Floquet code thereby ``misses" what is arguably the most interesting conceptual feature of the Kitaev 
honeycomb model -- the emergence of a gapless spin liquid, in which emergent Majorana fermions (coupled to a static
gauge field) exhibit a Dirac dispersion.
One reason for this is the strict temporal structure of the Floquet code, as it has recently been demonstrated that a {\it random}
measurement-only variant of the Kitaev model \cite{Vijay22Kitaev,Ippoliti22kitaev,Zhu23MajoranaLiquid} with no spatio/temporal ordering of XX, YY, and ZZ measurements
(but an underlying assignment of parity checks to bond types) 
leads to a Majorana liquid with characteristic $L \ln L$
entanglement structure (where $L$ is the linear dimension, proportional to the code distance, of the system).

In this manuscript, we pursue an alternative route to the rich physics of fractionalization and emergent Majorana liquids
by working directly with the honeycomb Floquet code, but tuning the {\it measurement strength} away from the strong, projective
measurements assumed in the code. Instead we explore the physics induced by weak measurements via injecting coherent 
noise into the system and moving the code away from Clifford stabilizer states~\cite{gottesman1997stabilizer}.
The phenomenology of such weak measurements on the quantum correlations of many-qubit states has recently been
theoretically explored in the context of GHZ states \cite{NishimoriCat, JYLee}, identifying a mapping to the Nishimori
physics of classical spin glass models \cite{Nishimori1981} which, crucially, allows to connect the measurement strength
to a temperature scale. 
We demonstrate that a similar connection can be made for the honeycomb Floquet code with weak measurements,
allowing us to induce physics akin to the finite-temperature phenomenology of the Kitaev model, in which the fractionalization
of spins and the formation of gauge ordering lead to two distinct thermal signatures~\cite{Nasu14mc,Nasu15kitaevfiniteT}.
At the same time, giving up strong 
measurements effectively moves the steady state of the modified Floquet code 
away from the corners of the corresponding ground-state phase diagram towards its center. Indeed, our numerical simulations
on finite system sizes indicate a pseudo-threshold at which the code breaks down due to flux proliferation, and the Majorana fermions escape the confinement of projective measurements forming a Majorana liquid -- the signature of which again is an $L \ln L$ entanglement structure observed in the entanglement negativity of the ensuing mixed state.


\begin{figure}[bt]
   \centering
   \includegraphics[width=\columnwidth]{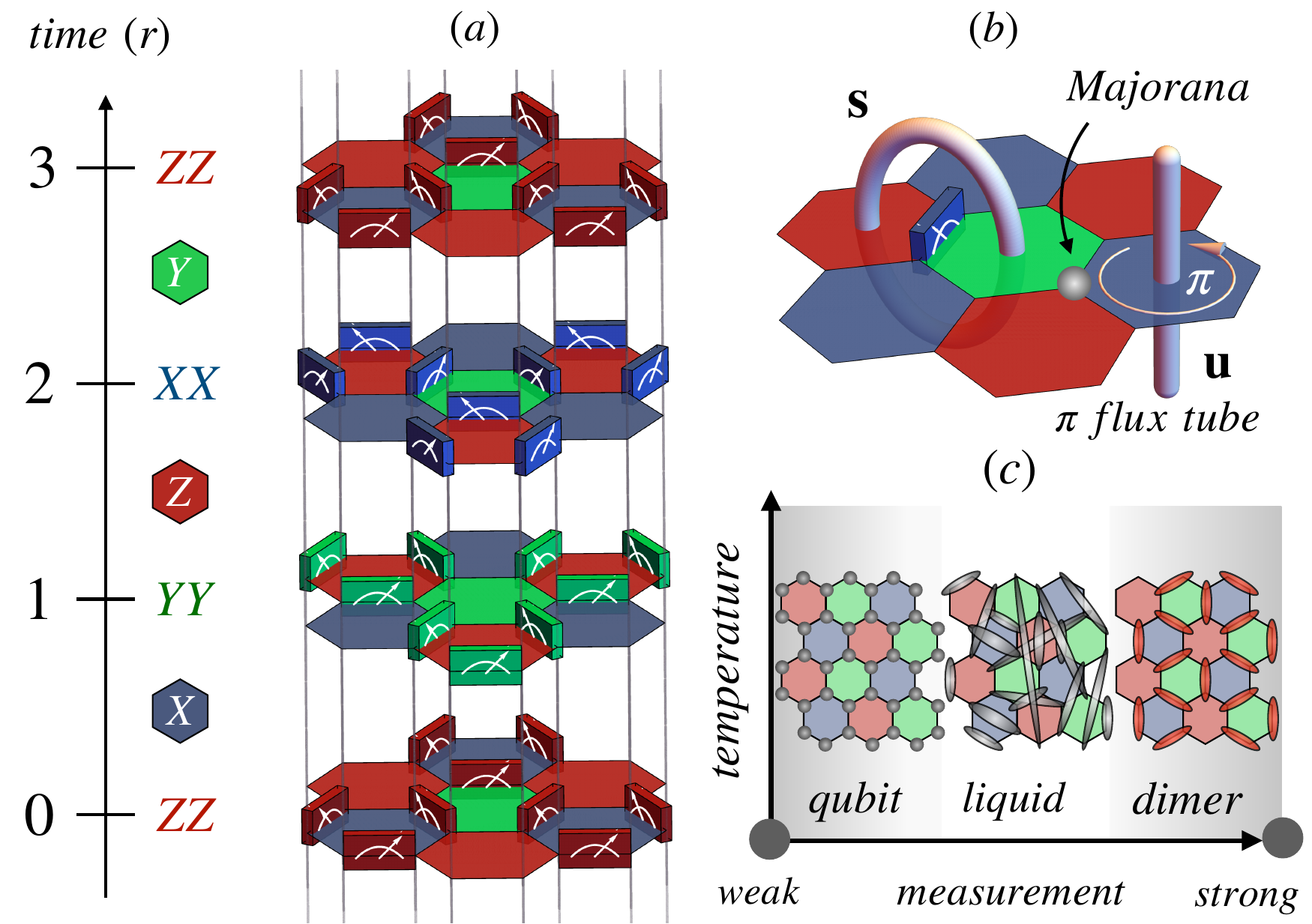} 
   \caption{{\bf Model}. 
   (a) {\bf Quantum circuit.} On a hexagon lattice, the plaquettes are tricolored into red (R), green (G), blue (B) islands, 
   bridged by complementary R, G, B bonds, respectively. 
   The Floquet circuit~\cite{Hastings22honeycomb} is composed of repeated, sequential 2-qubit parity measurements in a color dependent basis: 
   R $\to Z$, G $\to Y, $\ B $\to X$, respectively, 
   as visualized by each layer for one time step. Both the time and the spatial dimensions are period-3 translation invariant. 
   The hexagonal operators (e.g. $\hat{W} = \prod_{j\in\hexagon} Z_j$ for R) wrapped by two layers deduce 
   from the bond measurement outcomes. 
   (b)
  {\bf Fermion and flux.} The qubits fractionalize into Majorana fermions evolving in a Gaussian circuit under the background gauge flux tubes. A negative bond measurement outcome $s_{ij,r}=-1$ fluctuates a local flux loop. 
   (c) 
   {\bf Schematic phase diagram.}
   Increasing the circuit depth $r$ purifies the circuit-stabilized state, effectively lowering the relevant temperature scale to zero.
   It leads to an entanglement dichotomy of a long-range entangled Majorana liquid phase at intermediate measurement strength 
   and a short-range dimerized Majorana phase with toric code for strong measurements. 
   The schematics show typical configurations of the Majoranas, going from unpaired, localized modes over long-range pairs
   to tightly bound local dimers.   
   }
   \label{fig:protocol}
\end{figure}


{\it Protocol}.--
The Hastings and Haah Floquet code~\cite{Haah21honeycomb} consists of system qubits on the sites of a honeycomb lattice, which are subject to a time-periodic sequence of non-commuting parity measurements following a Kekul\'e, tricoloring of the bonds~\cite{Schmidt10honeycomb, Hastings22honeycomb}, as illustrated in Fig.~\ref{fig:protocol}(a). 
This translation invariant spacetime sequence is designed such that (after initialization) every measurement round allows to indirectly extract information about one of the three types of plaquette fluxes, the crucial ingredient to stabilize the three alternating toric code states. 
The two-qubit (weak) measurements can be implemented by introducing auxiliary qubits on every bond and coupling them to the system qubits via two-qubit unitaries~\cite{NishimoriCat, Chen23nishimori} (for details see SM~\cite{supplement}). 
The resultant non-unitary two-body Kraus operator (each brick in Fig.~\ref{fig:protocol}a) on the system qubits then reads
\begin{equation}
K_{ij}^\mu(s_{ij}) \equiv \exp\left(-\frac{\tau}{2}s_{ij}\sigma_i^\mu\sigma_j^\mu\right) / \sqrt{2 \cosh(\tau)}\ ,
\label{eq:KrausOp}
\end{equation}
where $\mu = x, y, z$, and $s=\pm 1$ is the measurement outcome of the auxiliary qubit, 
and $\tau\in[0,+\infty)$ characterizes the strength of measurement that is controlled by a unitary entangling gate parameter $t\in[0,\pi/4]$: 
$\tanh(\tau/2)=\tan(t)$ where $t = \pi/4$ corresponds to the strong, projective measurement limit.
After each measurement round, the auxiliary qubits are reset and recycled for next round, generating a dynamical sequence of measurement outcomes in the form of a spacetime resolved binary number, which we denote as $\mathbf{s}$. 
The product of Eq.~\eqref{eq:KrausOp} along such a recorded sequence defines a Kraus operator $K_\mathbf{s}$ as back-action, with spacetime random bond disorder, to the quantum state of system qubits. 
Due to the back-action, the quantum expectation value is (weakly or strongly) correlated with the classical readout. 
For example, the parity expectation can be found in SM~\cite{supplement}. 
Most importantly, every two consecutive rounds of measurements envelop a set of plaquettes (of the same color), where the plaquette operator $\hat{W} = \prod_{j\in\hexagon} \sigma_j^\mu$ is correlated with the product of the measurement outcomes $W_s = \prod_{ij\in\hexagon} s_{ij}$ around the plaquette (thus monitoring the plaquette flux), indicated in the interval between two time steps in Fig.~\ref{fig:protocol}(a). 
Without loss of generality we consider periodic spatial boundary conditions, initiate the state from a maximally mixed state, and run the protocol up to a (deep) circuit depth $r=L$~\footnote{We leave a discussion for the depth-3 protocol for free boundary conditions, initializing from pure product state to the supplementary material (SM).}. 


\begin{figure*}[tb] 
   \centering
   \includegraphics[width=\linewidth]{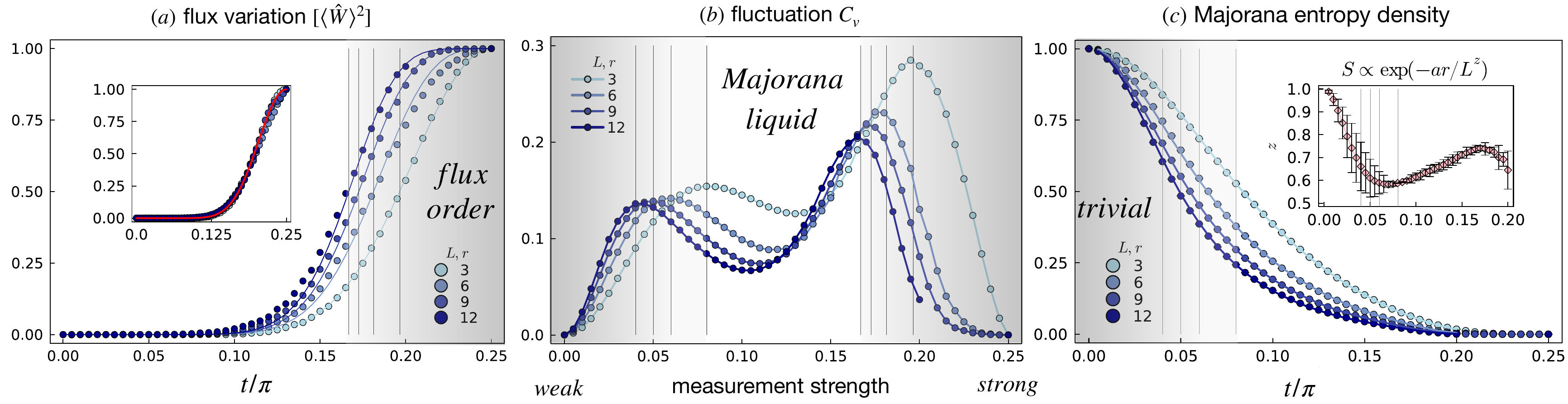} 
   \caption{
   {\bf Code stability and purification.}
   (a) {\bf Gauge flux observable.}
   Circles show $[\langle \hat{W}\rangle^2]$ that manifests the purification of fluxes. 
   The solid curves denote an analytic ansatz (see main text). 
   The vertical lines denote a ``pseudo-threshold" by taking the value of $[\langle \hat{W}\rangle^2]=1/2$, 
   terminating a gray shaded stable code regime.    
   Inset: 
   Rescaling collapses all data on top of $ \sin(2t)^{12}$ (red line).
   (b) {\bf Double peaks of fluctuation.} The two-peak structure signals the fractionalization of qubits into Majorana fermions and gauge fields. 
   The first peak signals the fluctuations of Majorana fermions, while the second peak captures the fluctuations of the gauge field. 
   The gray shades guide the eyes to the stable code zone (right panel) and a weakly monitored zone (left panel), 
   whereas the intermediate window indicates an emergent Majorana liquid.
   The gray vertical lines on the left zone denote the peak locations for each finite size; 
   the gray vertical lines on the right zone denote the pseudo-thresholds determined in (a) 
   which agree well with the peak of fluctuations. 
   (c)  {\bf Majorana entropy per site} in unit of $\ln 2$, where the circles denote Monte Carlo results 
   while solid lines denote a fitted scaling form $S/N = \exp\left(-a r/L^z\right)$. 
   Note that since time/circuit depth and system size are locked $r=L$, the tendency of the lines not only reflects the system-size dependence 
   but also the purification dynamics. 
   The latter clearly exhibits a system-size dependence in the weakly monitored regime, indicative of critical scaling. 
   Fits of the entropy density for each $t$ to a critical scaling form with a dynamical critical exponent $z$,
   which determines the purification timescale, leads to the results shown in the inset.
   Note that $z\to 1$ for weak measurements indicating Lorentz invariance, but drops below 1,i.e.\ faster than ballistic,
   for intermediate measurement strengths. 
   }
   \label{fig:W2}
\end{figure*}


{\it Majorana fermions and gauge field}.--
To discuss the many-body physics induced by this Floquet protocol, we resort to the parton language familiar from Kitaev's 
analytical solution of the Hamiltonian system~\cite{Kitaev2006}. As such, we note that like the quantum spins in the Hamiltonian language, 
the system qubits of the measurement protocol at hand can be mapped to a Majorana fermion coupled to a gauge field,
which remains subject to a gauge constraint~\cite{Hermanns2018, Trebst20kitaev3d}.
Here such a parton construction maps the (2+1)D quantum circuit to a {\it Gaussian fermion circuit}~\cite{DiVincenzo02freefermion, Jian22network, Fidkowski2021howdynamicalquantum}, 
coupled to a gauge flux. 
To be concrete, each Kraus operator is decomposed into a block diagonal form: $K_{\mathbf{s}} = \sum_{\mathbf{u}}K_{\mathbf{su}}\otimes \ket{\mathbf{u}}\bra{\mathbf{u}}$, where $u_{ij}=\pm 1$ is the internal Ising gauge field on the bonds, and $K_{\mathbf{su}}$ is a free-fermion evolution coupled to the {\it product} of $\mathbf{s}$ and $\mathbf{u}$. 
Conventionally, the Majorana fermion experiences a conserved static flux tube from $\mathbf{u}$ that stretches straight in time dimension~\cite{Kitaev2006}, but the measurement record $\mathbf{s}$ can fluctuate in time, creating local loop of flux tube $\mathbf{su}$ experienced by Majorana, schematically illustrated in Fig.~\ref{fig:protocol}(b). 
As a result, the post-measurement state of a given measurement record $\mathbf{s}$ is
$
\hat{\rho}_\mathbf{s} =\frac{1}{P(\mathbf{s})} \sum_{\mathbf{u}} p_{\mathbf{su}}\rho_{\mathbf{su}} \otimes \ket{\mathbf{u}}\bra{\mathbf{u}},
$
where $\mathbf{c}$ is the Majorana fermion on the system qubit sites. 
$\rho_{\mathbf{su}}$ is a normalized Gaussian fermion matrix describing a channel
\begin{equation}
\rho_{\mathbf{su}} := \frac{1}{p_{\mathbf{su}}} K_{\mathbf{su}}K_{\mathbf{su}}^\dag =\frac{1}{B p_{\mathbf{su}}}\exp\left(-\frac{\beta}{4} \mathbf{c} H_{\mathbf{su}} \mathbf{c}\right) \ ,
\label{eq:Ham}
\end{equation}
where $B=(2\cosh(\tau))^{(r+1)N/2}$ is a normalization constant, $\beta = r+1$ denotes the circuit-depth, and the channel probability is proportional to the Majorana partition function
\begin{equation}
p_{\mathbf{su}}= \text{Tr}\left(e^{-\frac{\beta}{4} \mathbf{c} H_{\mathbf{su}} \mathbf{c}} \right)/B\ ,
\label{eq:psu}
\end{equation}
where $H_{\mathbf{su}}$ is an $N$-by-$N$-dimensional single-fermion matrix. 
The probability of a given measurement record is obtained by summing over the internal gauge field $P(\mathbf{s}) = \sum_{\mathbf{u}} p_{\mathbf{su}}$.
Eq.~\eqref{eq:Ham} can be interpreted as an {\it effective Hamiltonian} $\frac{1}{4} \mathbf{c} H_{\mathbf{su}} \mathbf{c}$  at inverse temperature $\beta$, depending on the measurement strength $t\in[0,\pi/4]$. 
This effective Hamiltonian thereby relates the dynamically evolving, instantaneous state to a finite-temperature Gibbs state
\footnote{The eigen ``energies'' of this effective Hamiltonian encode the Lyapunov exponents that govern the dynamics~\cite{Pixley22MIPTCFT} (for numerics see SM~\cite{supplement}). 
In terms of Altland-Zirnbauer symmetry classes, this Hamiltonian belongs to the symmetry class BDI. As such, we note that the
Majorana liquid which we discuss in this manuscript is distinct from the ``thermal metal'' discussed in
conjunction with symmetry class D \cite{Laumann12thermalmetal} and the finite-temperature Kitaev Hamiltonian in the presence of a 
magnetic field \cite{Pachos19thermalmetal}.}.

Shifting attention from the qubit basis to the Majorana basis, we will dub the spacetime fluctuating $\mathbf{su}$ a {\it gauge trajectory}. The trajectory probability $p_{\mathbf{su}}$ is gauge invariant under a local gauge transformation: $c_j\to -c_j,\ u_{ij}\to -u_{ij}\ \forall i$, which does not change the flux configuration. 
Note that the ensemble of measurement records $\{ \mathbf{s} \}$ has an additional (0+1)-dimensional subsystem symmetry: 
at {\it all times} a flip for a fixed bond $s_{ij, r} \to -s_{ij, r}, \forall r$ leaves the probability $P(\mathbf{s})$ invariant, 
i.e. $P(\mathbf{s})=P(\mathbf{su})$. 

{\it Quantum Monte Carlo sampling}.--
The key observation is that each decomposed gauge trajectory has positive semi-definite probability Eq.~\eqref{eq:psu}, which
 can be polynomially computed by tracing out the fermions from the Gaussian fermion circuit~\cite{Calabrese10covariant, Zhu21quenchkitaev, supplement} and serves as a micro-step for a sign-problem free Monte Carlo sampling~\cite{Nasu15kitaevfiniteT}. 
Distinct from the finite-temperature Kitaev Hamiltonian, we here have two types of bond variables $\mathbf{s}$ and $\mathbf{u}$. Thus we need to perform a two-step nested Monte Carlo sampling: first sample an equilibrium ensemble of $\mathbf{s}$ (loops in Fig.~\ref{fig:protocol}(b)) via a Markov chain, and then, secondly, 
branch out from the equilibrium Markov chain to sample the time independent $\mathbf{u}$ (straight tubes in Fig.~\ref{fig:protocol}(b)) while fixing an $\mathbf{s}$ instance~\cite{supplement}. 
This numerically is quite costly, as the sampling space $\mathbf{u}$ consists of $3L^2$ bits while $\mathbf{s}$ consists of $L^2(L+1)$ bits.


{\it Two stage purification}.--
Due to the intrinsic randomness induced by measurements, a typical post-measurement state exhibits a random gauge flux configuration. 
Akin to an arbitrary toric code eigenstate, such a post-measurement state can be characterized by a non-linear~\cite{Preskill2003} (Edwards-Anderson) order parameter 
of the  flux observable
\begin{equation}
\begin{split}
[\langle \hat{W}\rangle^2]
=
\sum_{\mathbf{s}} P(\mathbf{s}) \langle \hat{W}\rangle_{\mathbf{s}}^2 
=
\sum_{\mathbf{s}, \mathbf{u}}  \frac{p_{\mathbf{s}} \cdot p_{\mathbf{su}}}{P(\mathbf{s})} \left(\prod_{l\in \hexagon} u_l \right)\ .
\end{split}
\label{eq:W2}
\end{equation}
Here we use $\langle \cdots\rangle$ to denote the quantum average (tracing out fermions and internal gauge field) and $[\cdots]$ to denote the measurement average. 
Note that Eq.~\eqref{eq:W2} involves {\it two} probability functions $p_{\mathbf{su}}$ and $p_{\mathbf{s}}$ expressing the expectation value of (i) pumping a static gauge field $\mathbf{u}$ onto (ii) the background of a gauge trajectory $\mathbf{s}$ favored by the Majorana fermions
\footnote{To decode the physical observables when running our protocol on a quantum device, one could use assistant classical computational resources to simulate the protocol under the {\it same} trajectory $\mathbf{s}$, to decode the quantum-classic cross-correlation~\cite{Garratt22, JYLee,Chen23nishimori}.}.
Projective measurements at $t=\pi/4$ yield $\langle \hat{W}\rangle = W_s =  \pm 1$
(the sign depending on the measurement outcome), a vanishing measurement average $[\langle \hat{W}\rangle] = 0$ with variance $[\langle \hat{W}\rangle^2]=1$.
From the Majorana fermion perspective, a zero {\it net} flux state is favored in this limit because of $\langle \hat{W}\rangle W_{\bf s}= +1$ (akin to the Lieb theorem for the corresponding
Hamiltonian ground state).
This implies that, for projective measurements ($t=\pi/4$), the flux induced by $\mathbf{s}$ is {\it screened} by the internal gauge field $\mathbf{u}$.

As shown in Fig.~\ref{fig:W2}(a) this flux order parameter remains close to 1 as we decrease the measurement strength, but eventually
drops to zero for intermediate measurement strength. 
This decaying pattern can be qualitatively explained by exponentially fast purification with a constant rate 
of  extracting information about the conserved gauge flux~\cite{Castelnovo2007, Vijay22Kitaev, Ippoliti22kitaev,Liu23coherent}.
We can define a proxy of the flux entropy, where the exponential decaying factor can be deduced 
\begin{equation*}
S_u:= -\log_2\frac{1+[\langle \hat{W}\rangle^2]}{2}\approx \left(-\log_2\frac{1+\sin(2t)^{12}}{2}\right)^{\frac{r+1}{4}} \ ,
\end{equation*}
which appears to capture the numerical behavior quite well, see the data collapse shown inset of Fig.~\ref{fig:W2}(a). 
Importantly, even when the flux purifies $\langle\hat{W}\rangle^2\to 1$ after sufficiently long times, for non-projective measurements $t<\pi/4$,
the internal gauge field $\bf u$ can no longer screen the fluctuating $\pi$ flux loops (encoded in the measurement outcomes $\mathbf{s}$).

We take the location for $[\langle \hat{W}\rangle^2]=50\%$ as an empirical pseudo-threshold for the crossover, at which the toric code dynamically stabilized by
the Floquet code breaks down~\cite{Castelnovo2007, NishimoriCat, Liu23coherent}. With increasing system size and number of measurement rounds, this pseudo-threshold slowly moves towards
smaller $t$, indicating that one can counterbalance the detrimental effect of weak measurements using deeper (and larger) codes.
%


To explore the state stabilized beyond this threshold, we rely on a similar diagnostic as for the finite-temperature Kitaev Hamiltonian 
where the {\it specific heat}, indicating the strengths of thermal energy fluctuations, is a useful measure to separate the temperature 
scales of gauge ordering (typically at temperatures
of around 1/100th of the coupling strength) and spin fractionalization (at a temperature corresponding to the coupling stength) \cite{Nasu15kitaevfiniteT,Trebst20kitaev3d}.
To diagnose fluctuations in the quantum circuit under weak measurement akin to the finite-temperature Hamiltonian, 
we look at the variation of the mean ``energy'' 
$E = \sum_{\mathbf{su}} p_{\mathbf{su}} E_{\mathbf{su}}$ of the effective Hamiltonian~\eqref{eq:Ham}.
Its variation is contributed by two parts~\cite{Nasu15kitaevfiniteT}
\begin{equation}
C_v =\frac{\beta^2}{N} \left([\langle E_{\mathbf{su}}^2\rangle ] - [\langle E_{\mathbf{su}}\rangle^2 ] + [\langle (\Delta E)_{\mathbf{su}}^2 \rangle] \right)\ ,
\end{equation}
where the first two terms quantify the fluctuations of energy among the internal gauge space, and the last term accounts for the energy fluctuations  in the Majorana fermion space. 
Numerical results are shown in Fig.~\ref{fig:W2}(b), where {\it two peaks} arise.
Moving $t<\pi/4$ out of the Clifford limit of strong measurements, the induced coherent noise first gives rise to a peak that capture the flux fluctuations at large $t$, which is followed by a peak reflecting the fractionalization of qubits into Majorana fermions at small $t$. 
To see the purification of the Majorana fermions more directly, we compute its entropy as $S_c = \beta (E - F)$, where $F$ the Majorana free energy averaged over $\mathbf{su}$. As shown in Fig.~\ref{fig:W2}(c), the purification of Majorana fermions (indicated by a decrease of this entropy) indeed sets in at the smaller measurement strength $t$ coinciding with the lower peak of the fluctuations.
Even for small circuits and depths we find a power-law decay with a long characteristic timescale $\propto L^z$, where the dynamical exponent $z$ varies with measurement strength at small $t$ and vanishes at large $t$, see the inset of Fig.~\ref{fig:W2}(c).

The two peak fluctuations, alongside the separation of purification of Majorana fermion and gauge flux are reminiscent of the Kitaev Hamiltonian at finite temperature~\cite{Nasu15kitaevfiniteT}. But distinct from the Hamiltonian where the Majorana fermions acquire coherence at a temperature that does {\it not} scale with the system size, the quantum circuit takes $L^z$ long time to purify the Majorana fermions. We will discuss in the following that this is in fact due to the formation of a Majorana liquid phase (at intermediate measurement strength), more entangled than the emerging Kitaev spin liquid~\cite{Kitaev2006}. \\

\begin{figure}[b] 
   \centering
   \includegraphics[width=\columnwidth]{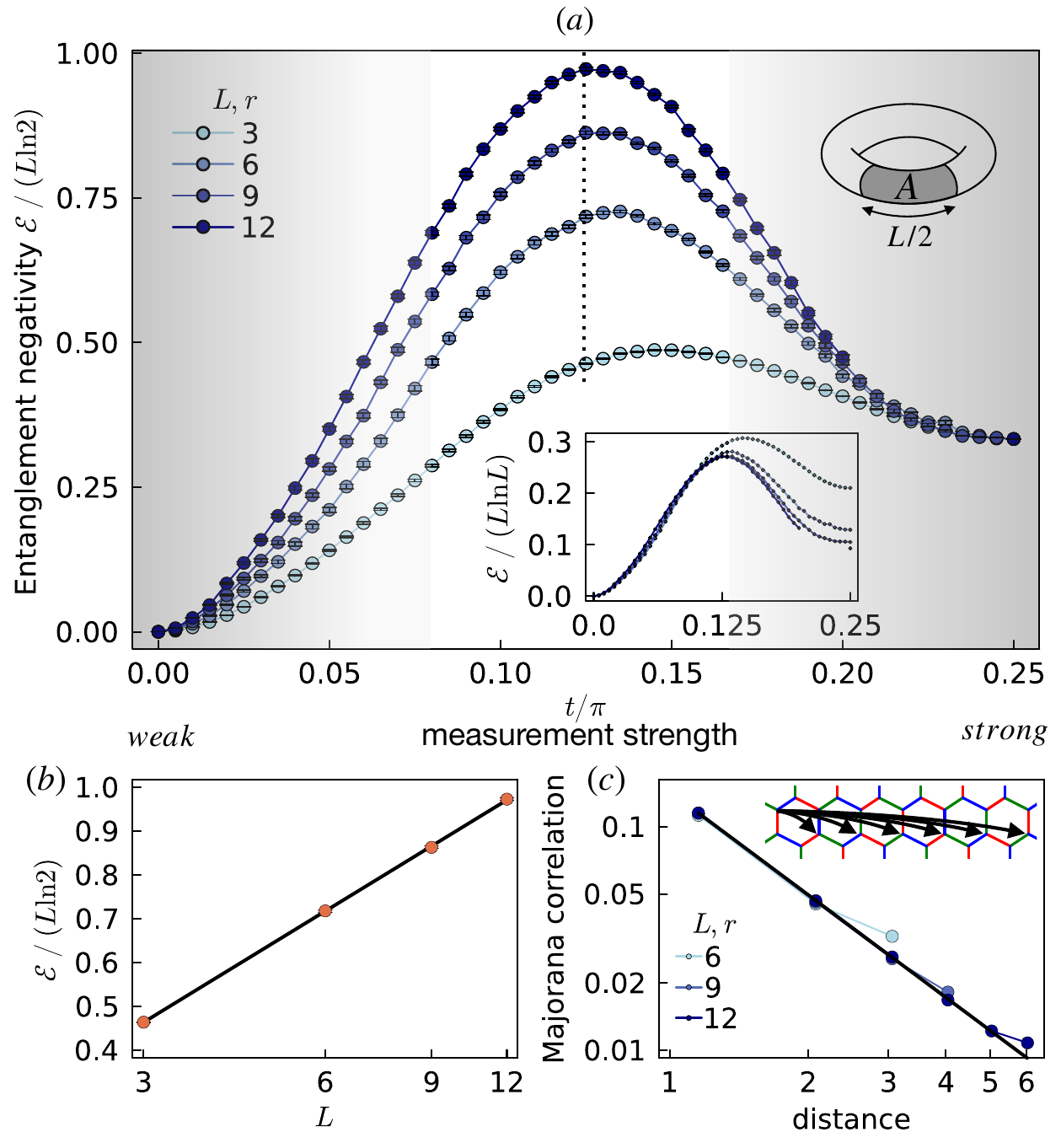}
   \caption{
   {\bf Majorana liquid}.
   (a) Entanglement negativity. 
   For intermediate measurement strengths, the entanglement negativity grows beyond area law with increasing system size, 
   following an $L\ln L$ scaling (b). 
   For strong measurements, the dynamically stabilized toric code state corresponding to a gapped dimerized Majorana state, 
   the entanglement negativity pivots to an area law and in the Clifford limit $t=\pi/4$ saturates at $\mathcal{E} = L(\ln2)/3$ (see main text).
   (b) The peak of negativity at intermediate $t=0.125\pi$ scales $\mathcal{E} \approx ( 0.18 L+ 1.10 L\ln L)(\ln2)/3$. 
   Additional plots of the entanglement negativity scaling for a wide range of measurement strengths $t$ 
   can be found in the SM. 
   (c) The real-space Majorana correlations  decay with a power-law at $t=0.125\pi$. 
   } 
   \label{fig:EntNeg}
\end{figure}


{\it Majorana liquid}.--
At intermediate measurement strengths, between the two peaks, there is an interesting entanglement dichotomy.
Whereas the fluxes proliferate akin to a high temperature toric code leading to only short-range entanglement~\cite{Grover20negativity},
the Majorana fermions give rise to long-range entanglement via the formation of a liquid state as we will now demonstrate.
Distilling the entanglement structure at intermediate measurement strength is a subtle issue owing to the mixed nature of the
purified state here. Instead of considering the entanglement entropy 
we turn to the the {\it entanglement negativity}~\cite{Vidal02negavitity}, which 
 is an effective probe of entanglement in a mixed state, as it allows to filter out the thermal contribution~\cite{Ryu19entnegT}
and can be readily  generalized to fermionic systems~\cite{Ryu17entneg, Eisler_2015, Ryu19entneg}. 
For its calculation, we bipartite the torus into two cylinders of equal lengths (see Fig.~\ref{fig:EntNeg}(a) inset) denoted as $A$ and its complement $\bar{A}$ and perform a partial time reversal $R_A$~\cite{Ryu17entneg} for the fermions in $A$, which transforms the fermionic density matrix to $\rho_{\mathbf{su}}^{R_A}$. The response under such a transformation indicates the entanglement between $A$ and $\bar{A}$, which is encoded in the negativity of its spectrum defined as
\begin{equation}
\mathcal{E} = \sum_{\mathbf{su}}p_{\mathbf{su}} \cdot \ln \norm{\rho_{\mathbf{su}}^{R_A}}_1\ ,
\end{equation}
where $\norm{\rho_{\mathbf{su}}^{R_A}}_1= \ln \text{Tr}\sqrt{\rho^{R_A}_{\mathbf{su}}\rho^{R_A \dag}_{\mathbf{su}}} $. Here we average over all the spacetime gauge configurations to obtain a typical negativity. 
As shown in Fig.~\ref{fig:EntNeg}(a), the entanglement of the typical fermion state grows monotonically with increasing circuit depth $r+1$, but non-monotonically upon varying the measurement strength $t$. Near the Clifford limit $t=\pi/4$, the negativity saturates to an area-law $\mathcal{E}\to L\ln(2)/3$. %
This limit can be easily understood, since the Majorana fermions form a ``valence bond crystal" of uncorrelated tightly bound dimers on one of the three bond types in every measurement round (Fig.~\ref{fig:protocol}(c)), with $2L/3$ dimers across the entanglement cut, each of which contributes half a bit of entanglement $\ln(2)/2$. 
By tuning $t<\pi/4$, the injected coherent error softens the Majorana dimers in each time step, which allows the Majorana fermion to propagate  to a longer distance in spacetime, purifying to a more entangled fermion state. 
In the intermediate regime around $t \sim \pi/8$, 
 the entanglement negativity clearly grows beyond area law, consistent with a $\mathcal{E}\propto L\ln L$ scaling with the system size as shown in Fig.~\ref{fig:EntNeg}(a,b) -- reminiscent to what is expected for a system with a Fermi surface~\cite{Wolf2006, Klich2006}. 
This is further supported by the observation of power-law decaying Majorana correlations in real space, shown in Fig.~\ref{fig:EntNeg}(c) for $t=\pi/8$.
Upon further decreasing the measurement strength, the prefactor of the $L\ln L$ scaling continuously decreases down to zero, similar to the  phenomenology of a Fermi surface shrinking to a Dirac point. We caution, however, that it takes deeper circuits of depth $r\sim L/t$ than what we have been able to simulate to fully purify in the scaling limit $t\ll 1$. 
We expect a Majorana entanglement phase transition 
(not necessarily at the same location as the pseudo-threshold defined for the flux ordering transition
\footnote{Note that there is yet another threshold to distinguish here: The pseudo-threshold of the flux ordering (see Fig.~\ref{fig:W2}(a))
and the Majorana entanglement phase transition bound a {\it decoding} transition \cite{Garratt22,JYLee} 
up to which one can decode the quantum memory
stabilized by the Floquet code in the presence of coherent noise (weak measurements).
}) 
that separates the $L\ln L$ entangled phase in the intermediate measurement regime from the area law phase in the strong measurement regime. This phase transition could be related to the physics of 3D Anderson localization transitions~\cite{Mirlin08rmp}, as the Gaussian fermionic circuits can be mapped to a 3D Chalker-Coddington network model~\cite{Jian22network} with {\it correlated} Born disorder~\cite{Nahum23measurefreefermion, Jian23measurefreefermion}, a generalization of the 2D random bond Ising model~\cite{Read2001, Chalker2002, NishimoriCat}. We leave it as a future study to elucidate the nature of the fermionic entanglement transition~\cite{Mirlin23measure2D, Buchhold23measure2D, Jian23measurefreefermion, Nahum23measurefreefermion}. 


{\it Discussion}.-- 
Let us close with a broader discussion of the conceptual connection of the honeycomb Floquet code and the Kitaev spin liquid (KSL). 
The two systems can be distinguished by their intrinsic {\it frustration}, a measure to what extent a state satisfies every local
generator of its underlying Hamiltonian or circuit, respectively.
While for the Kitaev spin liquid  frustration is unavoidable, resulting in long-range entanglement,
the (instantaneous) toric code state dynamically stabilized by the Floquet code is frustration-free in essence.
It is the {\it weak} measurement in the modified Floquet code that unlocks its frustration~\cite{NishimoriCat}, manifesting itself in the proliferation of  gauge fluxes that  interfere with the Majoranas. As a consequence, the Majorana fermions following {\it typical} gauge trajectories can reach a highly entangled liquid state with $L\ln L$ entanglement scaling. 
This state with its super-area-law entanglement turns out to be more entangled than the KSL with its area-law entanglement~\cite{Yao10KitaevEntanglement, Chen15entanglement, supplement} (independent of temperature).
If one wants to recover the KSL in a circuit, one would have to post-select the clean trajectory $\mathbf{s}=+1$ 
(with ordered or disordered $\mathbf{u}$ for zero or finite temperature, respectively) from the ensemble
of trajectories, i.e.\ the KSL is in fact a tiny subset of the ensemble of states out of the weak measurement Floquet protocol \footnote{If starting from maximally mixed state, one needs $rt\gg 10^2$ in order to lower the temperature down to $10^{-2}$ to purify the flux into the state ~\cite{Nasu15kitaevfiniteT} described by a Majorana Dirac fermion coupled to zero flux; however, this $10^2$ prefactor can be eased by using projective measurements at the beginning to pin the flux at constant depth.}.

In a similar vein, we can distinguish the emergent Majorana liquid in the weak measurement Floquet protocol from the one observed 
in the random projective measurement-only variant of the Kitaev model \cite{Nahum20freefermion, Vijay22Kitaev,Ippoliti22kitaev,Zhu23MajoranaLiquid} 
(with no spatiotemporal ordering of XX, YY, and ZZ measurements). Despite their similarity of an $L \ln L$ entanglement
scaling, they show very different levels of coherence and frustration. While the latter uses randomized Clifford measurements, which allow for
the teleportation of Majoranas (leading to long-range entanglement), the Majoranas are not truly {\it itinerant} but {\it localized} in a long-range valence bond crystal, this being frustration-free in disguise. 
In contrast, in going beyond the Clifford limit the weak measurement Floquet protocol allows long-range Majorana dimers to superpose 
as in a resonating valence bond liquid~\cite{Zhou17reviewQSL, Balents16reviewQSL}, typically a highly frustrated scenario. 

These two distinctions above show that the concepts of frustration and (non-)Cliffordness are deeply connected to one another, 
and with the latter also intrinsically linked to the concept of magic~\cite{Hamma22magic, Gullans23magic},
this points to a unifying organizing principle for entangled states of matter.\\

{\it Acknowledgements}.--
The authors would like to thank 
	Ruben Verresen, 
	Michael Buchhold,
	Ciarán Hickey, 
	Nathanan Tantivasadakarn,
	Achim Rosch, 
	Henning Shomerus, 
	Michael Gullans, 
	and
	Dominik Hangleiter
for helpful discussions. 
Our work was partially funded by the Deutsche Forschungsgemeinschaft under Germany's Excellence Strategy -- Cluster of Excellence Matter and Light for Quantum Computing (ML4Q) EXC 2004/1 -- 390534769 and within the CRC network TR 183 (Project Grant No. 277101999) as part of subprojects A04 and B01. 
The numerical simulations were performed on the JUWELS cluster at the Forschungszentrum Juelich. \\

{\it Data availability}.-- The numerical data shown in the figures is available on Zenodo~\cite{zenodo_qubitfrac}.

\bibliography{measurements}

\clearpage

\begin{center}
\large{SUPPLEMENTAL MATERIAL}
\end{center}

\tableofcontents

\section{Qubit based analytics}

\subsection{Circuit building blocks and Kraus operator}

\begin{figure}[htbp]    
	\centering
   \includegraphics[width=.8\columnwidth]{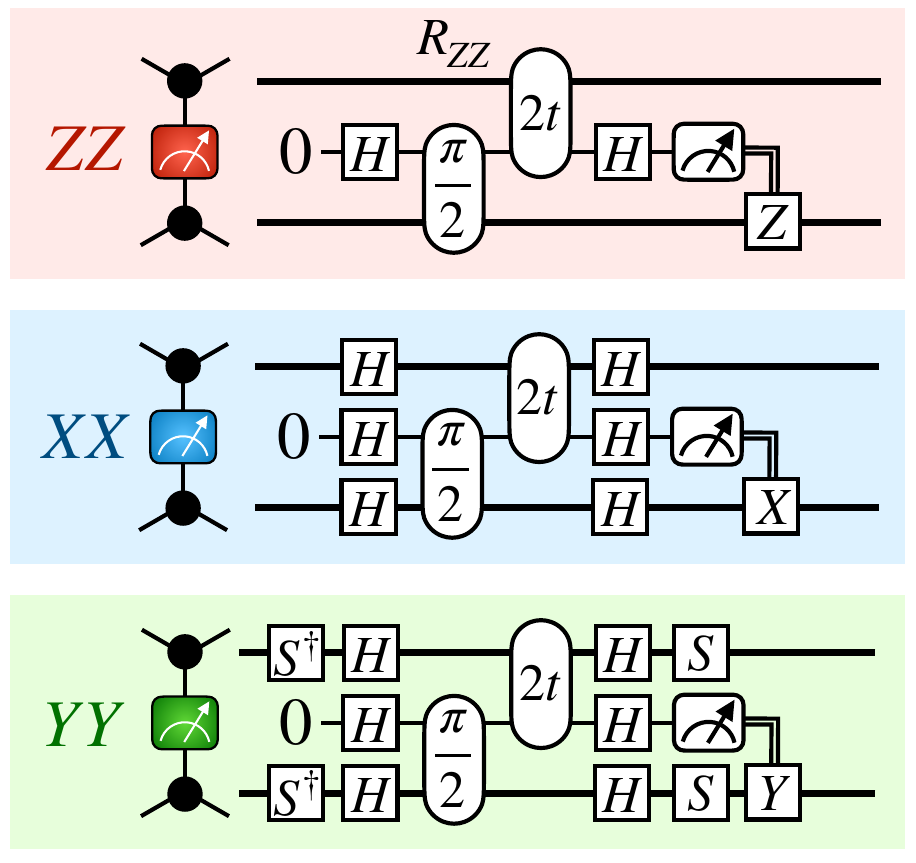} 
   \caption{{\bf Building blocks of weak parity measurement}, for measurement of $ZZ$ (in red), $YY$ (in green), $XX$ (in blue), respectively. The two-body entangling gates are $R_{ZZ}(\alpha) = e^{-i \frac{\alpha}{2} ZZ}$, sandwiched by one-body Hadamard gate $H$ or phase gate $S=\sqrt{Z}$ that rotates the Pauli basis. $t=\pi/4$ realizes the Hastings-Haah honeycomb Floquet code, which belongs to the Clifford circuit. Deviation from $\pi/4$ injects coherent error to the circuit. 
   The last single-Pauli operation in each block is conditioned upon negative measurement outcome, which can be pulled through the entire circuit till the final step and be post-processed on software level, similar with the treatment in measurement-based quantum computation. }
   \label{fig:circuit}
\end{figure}

Let us start with an in-depth description of how the two-qubit $XX$, $YY$, or $ZZ$ parity checks can be implemented
using an auxiliary qubit and single-qubit measurements. This is summarized in Fig.~\ref{fig:circuit}.

In mathematical terms, the weak measurement of interest in the manuscript at hand can be expressed by a Kraus operator induced by measuring out the ancilla qubit after coupling it to the two adjacent site qubit, followed by a conditional Pauli operator (realized in Fig.~\ref{fig:circuit})
\begin{equation}
\begin{split}
M_s
=&(i\sigma_B^\mu)^{\frac{1-s}{2}}\bra{ s^x\equiv s} e^{-i s^z \otimes (t \sigma_A^\mu+\frac{\pi}{4}\sigma_B^\mu)} |+\rangle\\
=&\frac{\cos t}{\sqrt{2}}
\begin{cases}
1- \tan t  \sigma_A^\mu\sigma_B^\mu, & s=+1\\
 1+ \tan t \sigma_A^\mu\sigma_B^\mu, & s=-1\\
\end{cases}\\
 =& \frac{1}{\sqrt{2\cosh \tau}}e^{-\frac{1}{2} \tau s\sigma_A^\mu\sigma_B^\mu}\ ,
\end{split}
\end{equation}
where $\tan t = \tanh\frac{\tau}{2}$ parametrizes the measurement strength, with the strong, Clifford limit corresponding to $t = \pi/4$.

\subsection{Parity measurements}

\begin{figure}[thb] 
   \centering
   \includegraphics[width=\columnwidth]{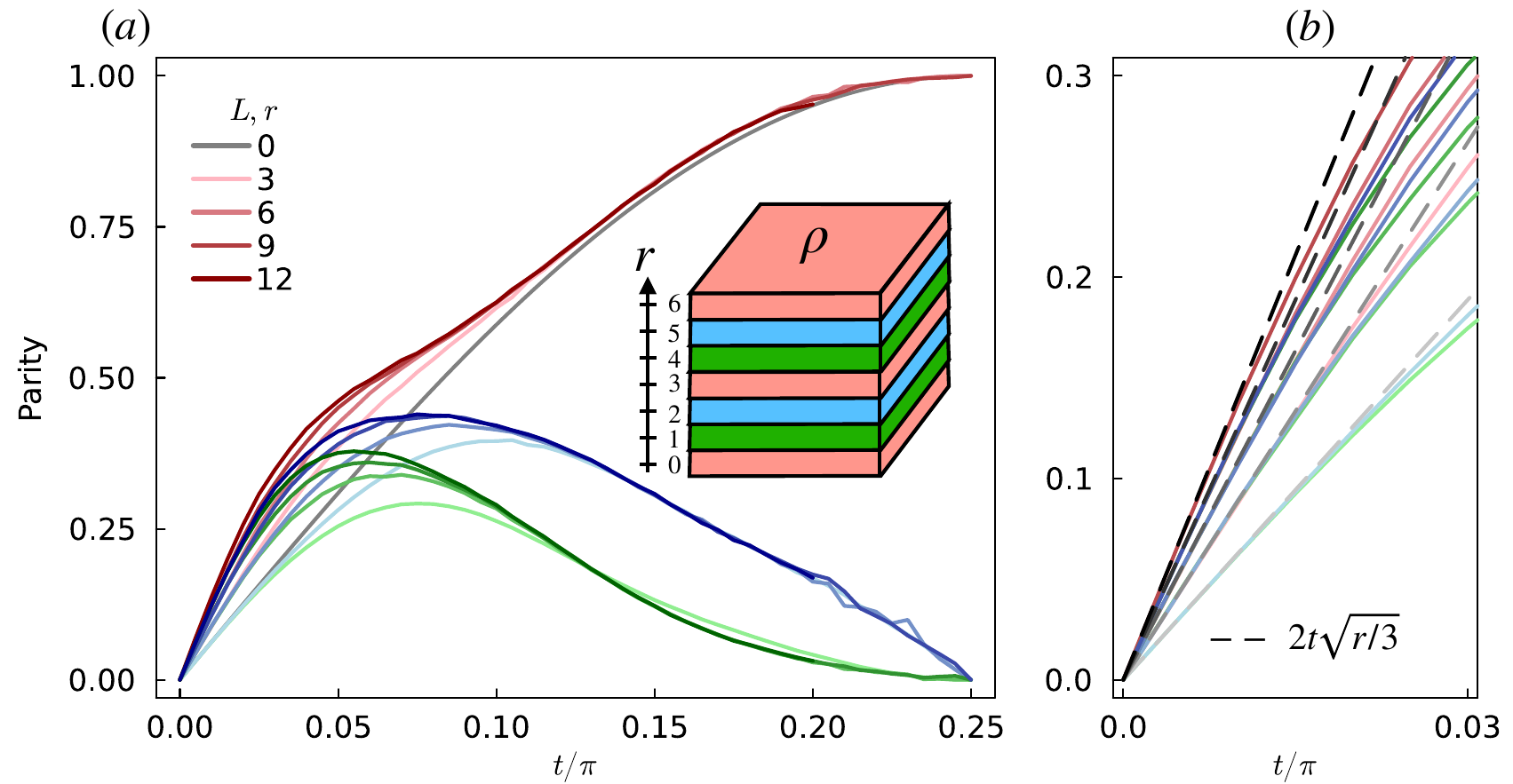} 
   \caption{
   {\bf Local two-qubit parity expectation} for the final state for the corresponding R, G, B bonds, respectively. 
   (a)The inset shows the schematic $2+1D$ spacetime evolution bulk composed of non-commuting tri-coloured layers, where the final state is a slab of the temporal boundary, that preserves 3-fold dimer rotation symmetry at $t\ll 1$ but explicitly breaks it at $t\to \pi/4$. The Clifford point $t=\pi/4$ can be analytically solved: $G=B=0, R=1$. 
   The measurement average is performed over the squared of the expectation to get rid of the random sign.
   Since R bonds are the ones being finally measured, they have strongest expectation $ZZ \to \pm 1$ when it approaches the Clifford limit $t\to \pi/4$, while G and B bonds have been suppressed by the latest anti-commuting measurements. 
   The black solid line is the quantum-classical cross-correlation for the latest measurement $[\langle \sigma_i^\mu\sigma_j^\mu \rangle \cdot s_{ij}] = \sin(2t)$ that lower-bounds the square root of EA dimer correlation. The deviation reflects the trajectory dependence of the $R$ bonds, which have been measured more than one times.  
   (b) A zoom in view of the weakly monitored regime $t\ll 1$ where the observables converge well to the analytic expression $2t\sqrt{r/3}$ (gray dashed lines). Note that $R(r)\sim 2t\sqrt{r/3+1}$ because it lies at the temporal boundary. Thus $R$ being measured in the previous period agrees with the $G$ and $B$ in the current round, because they are all in the bulk away from the temporal surface. 
   }
   \label{fig:kekuledimer}
\end{figure}

To see the impact of tunable, weak measurement, we start by computing the root mean square of the {\it two-qubit parity} operator that is being measured: $[\langle \sigma_i^\mu \sigma_j^\mu\rangle^2]^{1/2}$, as shown in Fig.~\ref{fig:kekuledimer}(a). Since we restrict ourselves to $r+1$ rounds of measurements and $r\mod 3 =0$, the R bonds are measured at latest, which becomes strongest and $\to 1$ when $t\to \pi/4$. In contrast, the G,B bonds anticommute with R bonds and are thus be overwritten and suppressed, becoming exactly zero in the Clifford limit $t\to \pi/4$. We can derive that the correlation between the quantum expectation $\langle Z_i Z_j\rangle_s$ and that of the last measurement outcome $s_{ij, r}$ follows an analytic expression: $[\langle Z_i Z_j \rangle_s \cdot s]=\sin(2t)$. Due to Cauchy-Schwarz inequality, it lower-bounds the Edwards-Anderson (EA) correlation $[\langle Z_i Z_j \rangle_s \cdot s]\leq [\langle Z_i Z_j \rangle^2]^{1/2}$, which is saturated only at the Clifford limit where the measurement outcome is equal weight distributed. 
In Fig.~\ref{fig:kekuledimer}(b) we compare the numerical results to the analytic expectation which is essentially repeated measurements for a single effective qubit. 

Here we derive an approximate description of the bond measurement ensembles at weak monitored regime $t\ll 1$, for repeated weak measurements for a single qubit. 
Consider a single Pauli string operator $O$ that satisfies $O^2=1$, then we have a 0+1-dimensional measurement outcome trajectory $\mathbf{s} :=\{s_n=\pm 1\}$, $n=1,\cdots,r$. Initializing from a maximally mixed state, after $r$ times of measurement, the post-measurement state is
\begin{equation}
\rho_{\mathbf{s}} \propto e^{\tau s_{tot} O},
\end{equation}
with normalized probability 
\begin{equation}
p_{\mathbf{s}} = \frac{\cosh(\tau s_{tot})}{(2\cosh\tau)^r},
\end{equation}
where $s_{tot}\equiv \sum_{n=1}^r s_n$ is the sum of the $r$ rounds of measurement outcomes. 
The autocorrelation of the classical readout, between any two times, is
\begin{equation}
[s_ms_n] =1- (\cosh\tau)^{-2},
\end{equation}
which means (i) they are long-range correlated in time axis; (ii) their correlation increases monotonically with increasing measurement strength and saturates at strong measurement limit. (Bear in mind that $\tanh\frac{\tau}{2} = \tan t$.) The correlation between the observable and the latest measurement outcome is always
\begin{equation}
[\langle O\rangle s_r] = \tanh\tau,
\end{equation}
which is related to the measurement strength but not the time because the histories are washed out. 
And the second moment correlation expresses the trajectory rigidity and depends on the depth
\begin{equation}
\begin{split}
[\langle O\rangle^2] &= \sum_{s_{tot}}\tanh(\tau s_{tot})^2 P(s_{tot}) \ .
\end{split}
\end{equation}
When the time dimension is traced out, the distribution of $s_{tot}$ is {\it bimodal} that originates from a superposition of the tails of two separated Gaussian distributions peaked at $\pm r\tau$ with variance $r$
\begin{equation}
\begin{split}
P(s_{tot}) &= \frac{1}{(2\cosh\tau)^r}\cosh(\tau s_{tot})\binom{r}{\frac{r+s_{tot}}{2}}\\
&\approx  \frac{1}{(\cosh\tau)^r \sqrt{2\pi r}} \cosh(\tau s_{tot}) e^{- \frac{1 }{2r}s_{tot}^2}\\
&\propto  
\left(
e^{-\frac{1}{2r}(s_{tot}+r\tau)^2}+ e^{-\frac{1}{2r}(s_{tot}-r\tau)^2}
\right) \ ,
\end{split}
\end{equation}
subjected to the constraint $-r\leq s_{tot} \leq +r$. 
For weak monitored regime $t\ll 1$, one can deduce that $[\langle O\rangle^2]\approx \tau^2 \cdot \text{var}(s_{tot})$. $P(s_{tot})$ forms an approximate Gaussian distribution with variance $r$. Consequently, $[\langle \sigma^\mu\sigma^\mu\rangle^2]\approx 2t\sqrt{r/3}$ for $t\ll 1$, that agrees with the numerical calculation in Fig.~\ref{fig:kekuledimer}b. 

\subsection{Correlation between quantum and classical flux}

Consider a single plaquette surrounded by two rounds of measurement, one can trace out the Majorana fermions $\mathbf{c}$ (or equivalently all open Pauli strings) leaving a reduced density matrix for the flux (closed loop configurations)
\begin{equation}
\text{Tr}_\mathbf{c}\rho_{\mathbf{s}} \propto 
1+(\tanh\tau)^6 \hat{W} W_{s} \ .
\end{equation}
Consequently,
\begin{equation}
[\langle \hat{W}\rangle W_s]=(\tanh\tau)^6
\label{eq:WWs}
\end{equation}
and $[\langle \hat{W}\rangle]=[W_s]=0$. Eq.~\eqref{eq:WWs} applies to more general case, as long as $W_s$ is the latest measurement readout, which basically expresses the instantaneous correlation between the measured quantum observable and the classical readout. We use this exact result as a sanity check for the Monte Carlo samples. For $L=12$ and $t>0.2\pi$, the Monte Carlo samples using simple update up to 2000 sweeps deviate from this analytic expression and are thus discarded. 

\begin{figure}[t] 
   \centering
   \includegraphics[width=.7\columnwidth]{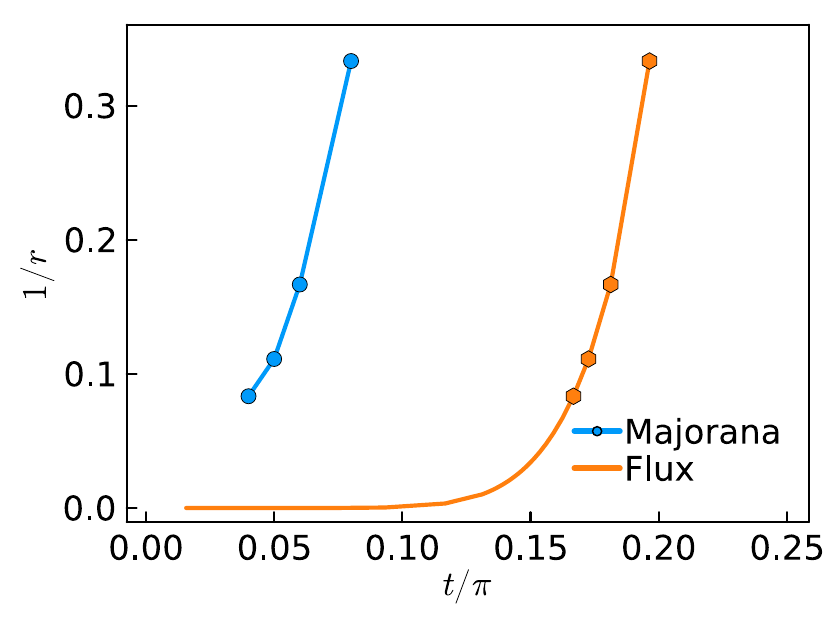} 
   \caption{{\bf Crossover threshold $t_c$ scaling} with circuit depth $r$, associated with Majorana purification (blue) and flux purification (orange), respectively. The blue circle is extracted from the peak locations of Majorana fluctuation $C_v$; the orange data is extracted from the analytic ansatz for exponential purification. }
   \label{fig:crossoverscaling}
\end{figure}

\subsection{Crossover threshold scaling}
A visualization of the scaling of the crossover $t_c$ is shown in Fig.~\ref{fig:crossoverscaling}. Even though the flux purifies exponentially fast (with exponential decaying entropy with circuit depth), the absolute value of $t_c$ appears to decay slow when increasing depth. 

\section{Fermion based statistical mechanics}

\begin{figure}[b]
   \centering
   \includegraphics[width=.6\columnwidth]{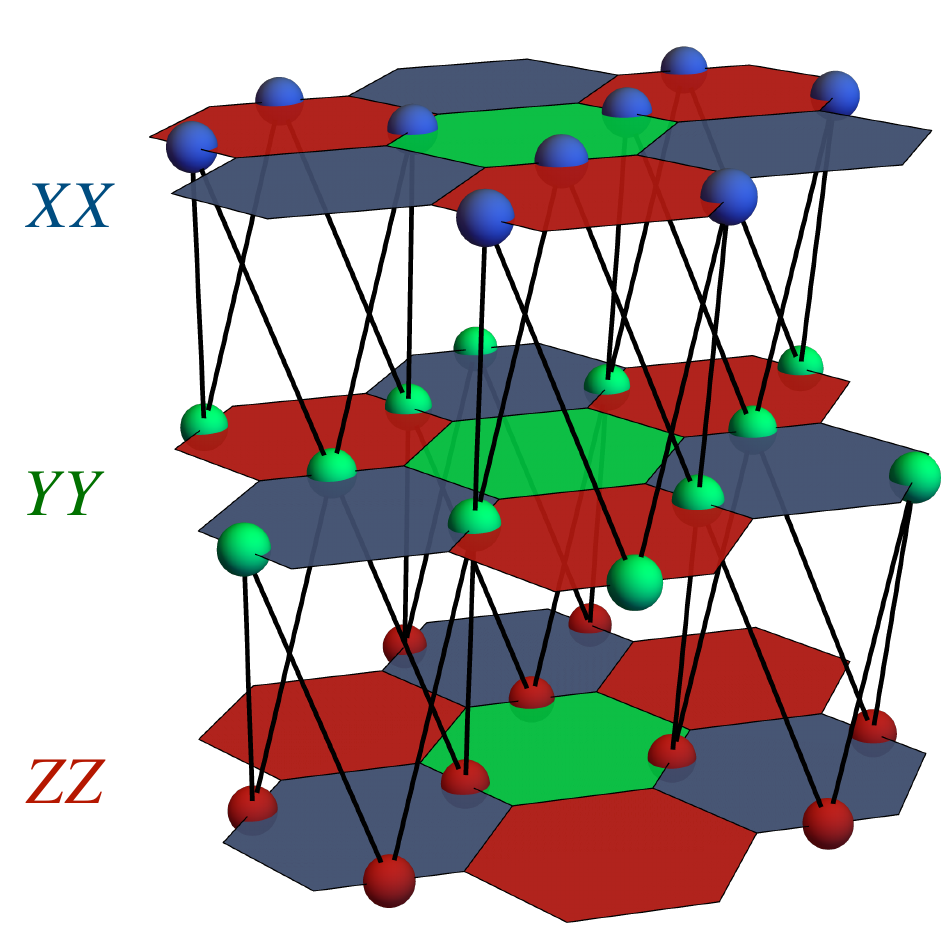} 
   \caption{
   {\bf The 3D free-fermion network model}. 
   The effective Chalker-Coddington network model in a given gauge trajectory $\mathbf{su}$. Each node originates from one two-qubit gate in (a) and is mapped to one scattering node with 4 legs, corresponding to a transfer matrix $e^{\tau su\sigma^y}$ where $\sigma^y$ denotes a Pauli matrix that should not be confused with the physical qubit. 
   }
   \label{fig:circuitnetwork}
\end{figure}
\subsection{Gaussian fermion evolution}

Under a fixed gauge trajectory $\mathbf{su}$, the quantum circuit reduces to a Gaussian fermion circuit. As shown in Fig.~\ref{fig:circuitnetwork}, each gate reduces to a scattering node with four legs representing the incoming and outgoing free fermion modes, as in a Chalker-Coddington network model. 
Each node is described by a transfer matrix:
\begin{equation}
\begin{split}
e^{\tau su \sigma^y} 
= e^{\ln(2\cosh(\tau))}\frac{1 +su \tanh(\tau)\sigma^y}{2}
\end{split}
\end{equation}
that linearly maps~\cite{Kitaev2006} the Majorana fermion operator under time evolution. Note that the measurement-induced binary disorder enters as a sign factor into the node, similar to the network model for 2D random bond Ising model~\cite{Chalker2002}. 
Collecting all the nodes in one circuit layer, we get an $N$-by-$N$-dimensional transfer matrix for the entire system, where each node contributes a 2-by-2 block. We denote the transfer matrix of layer $n$ by $e^{A_n}\equiv e^{\tilde{A}_n+\ln(2\cosh(\tau))}$, where the rescaled exponent $\tilde{A}_n$ has its spectrum upper-bounded by $0$. 
Collecting all the layers for the entire circuit, the Gaussian fermion operator for the complete trajectory 
\begin{equation}
\rho_{\mathbf{su}} \propto \left(\prod_{n=0}^{r}e^{\frac{1}{4} \mathbf{c} A_n\mathbf{c}} \right)\times h.c.= e^{-\frac{\beta}{4} \mathbf{c}H_{\mathbf{su}}\mathbf{c}}
\end{equation}  
can be computed by using the Baker Hausdorff formula:
\begin{equation}
H_{\mathbf{su}}=-\frac{1}{\beta}\ln  \left(e^{A_r} \cdots e^{A_0} \times e^{A_0} \cdots e^{A_r}\right) \ .
\label{eq:effHam}
\end{equation}
Then the hermitian and antisymmetric Majorana covariant matrix follows as
\begin{equation}
i\Gamma_{\mathbf{su}} 
\equiv \frac{i}{2p_{\mathbf{su}}} \text{Tr} \left(\rho_{\mathbf{su}} [c, c^T] \right) 
 =
- i\tanh\frac{\beta H_{\mathbf{su}}}{2}\ .
\end{equation}
The internal energy is 
\begin{equation}
E_{\mathbf{su}} = \frac{1}{4p_{\mathbf{su}}} \text{Tr} [\rho_{\mathbf{su}}(\mathbf{c} H_{\mathbf{su}} \mathbf{c})]
=
\frac{1}{4}\text{tr}\left(H_{\mathbf{su}} \Gamma_{\mathbf{su}}\right)
 \ ,
\end{equation}
where we use $\text{tr}(\cdots)$ to denote single-particle trace in $N$-by-$N$-dimensional space. 
The energy variance in the Majorana space can be obtained by Wick decomposition. 
By diagonalizing $H_{\mathbf{su}}$ for its eigenvalues $\{\pm\epsilon\}$, we have the quantities expressed by the single-mode eigen-energies: 
\begin{equation}
\begin{split}
F_{\mathbf{su}}&=-\frac{1}{2\beta}\sum_{\epsilon\leq 0} \ln\left(2(1+\cosh(\beta\epsilon))\right)\ ,\\
E_{\mathbf{su}}&=-\frac{1}{2}\sum_{\epsilon\leq 0}\left(\epsilon\tanh\frac{\beta\epsilon}{2}\right)\ , \\
\langle (\Delta H)^2\rangle_{\mathbf{su}} &= \frac{1}{4}\sum_{\epsilon\leq 0}\left(\epsilon^2\left(1-(\tanh\frac{\beta\epsilon}{2})^2\right)\right)\ .
\end{split}
\end{equation}
By linearly averaging them with weight $p_{\mathbf{su}}$, we obtain the average Majorana free energy $F$, internal energy $E$, energy variation, respectively. The average Majorana entropy can be calculated by $S_c = \beta(E - F)$, which should be distinguished from the entropy of flux or trajectory.

{\it Numerical computation}.--
In practice we start from the middle of the chain in Eq.~\eqref{eq:effHam} and iteratively absorb the rescaled transfer matrix:
$
e^{-\beta\tilde{H}_n} = e^{\tilde{A}_n} e^{-\beta\tilde{H}_{n-1}} e^{\tilde{A}_n}
$, by starting with $\tilde{H}_{-1}=0$. 
The final matrix $e^{-\beta\tilde{H}_r}$ also has its spectrum bounded within the unit-circle. Taking the logarithm for the eigenvalues, we can retrieve the exponent $H_r = \tilde{H}_r -2\ln(2\cosh(\tau))$. With the particle-hole symmetry the spectrum should be symmetric around $0$, and thus we only record the negative branch that dominates the dynamics and suffers less from the rounding error. 
An alternative way is to keep track of the covariant matrix $\Gamma_n:= -\tanh(\beta H_n/2)$ evolution:
$
\Gamma_n = \Gamma_{A_n} \times \Gamma_{n-1} \times \Gamma_{A_n}
$, where $\Gamma_{A_n} = \tanh(A_n/2) = \tanh(\tau/2) su \sigma^y$ is the covariant matrix for each step evolution, and the composition operation $x \times y $ here is not matrix product but rather $x\times y:=1 - (1-y) (1+xy)^{-1} (1-x)$~\cite{Calabrese10covariant}. 
Note that such fermion evolution computation suffers from numerical singularity when $t\to \pi/4$ and for deeper circuit, because of diverging $\beta \epsilon$. This can overcome by iteratively rescaling the accumulated transfer matrix as commonly done in a localization problem.

\subsection{Lyapunov exponent and the trajectory entropy}

\begin{figure}[h!] 
   \centering
   \includegraphics[width=\columnwidth]{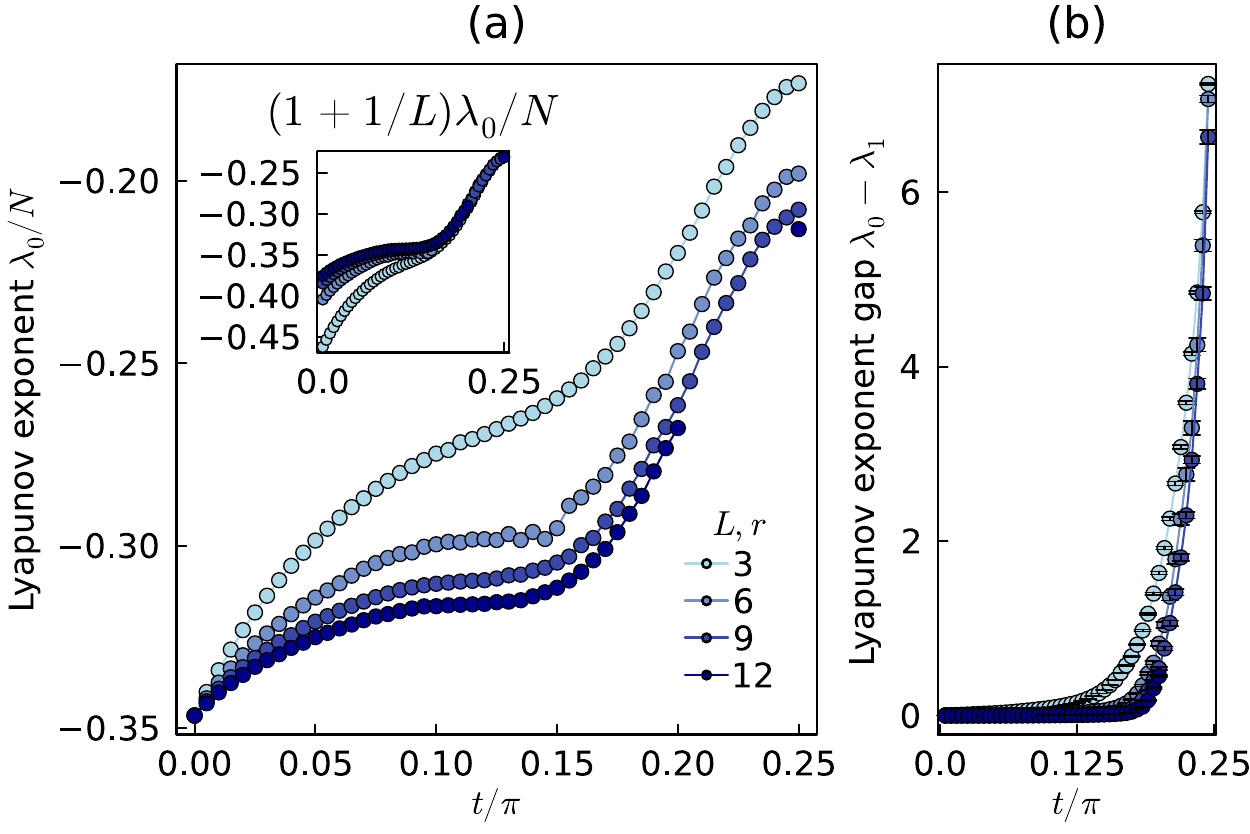} 
   \includegraphics[width=\columnwidth]{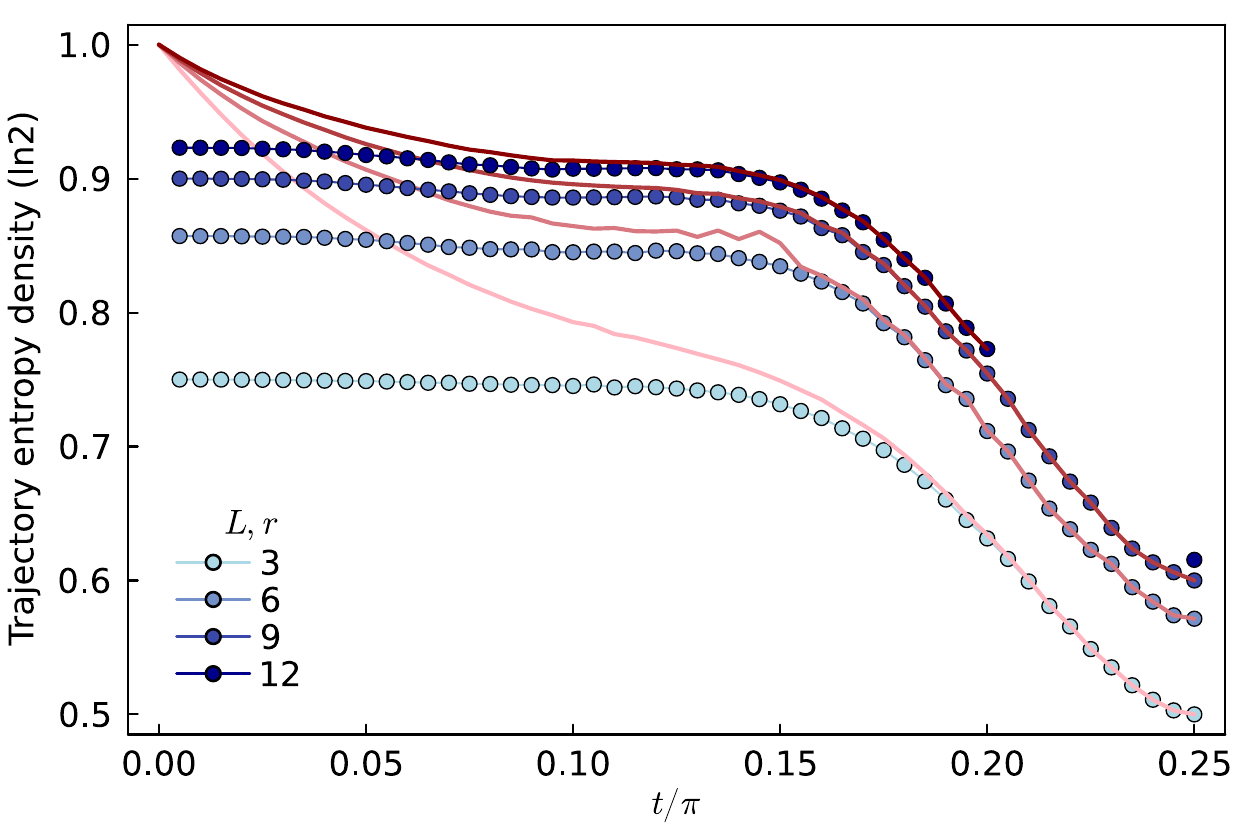} 
   \caption{
   {\bf Lyapunov exponent and entropy of net gauge trajectory}. 
   (a) The leading Lyapunov exponent per site, equivalent to the ground state energy density. At $t=0$, $\lambda_0/N = -\ln(2)/2$, while at $t=\pi/4$ it is solved to be $\lambda_0/N = -\ln(2)/(3(1+1/L))$. 
   (b)The gap between the leading Lyapunov exponent $\lambda_0$ and the subleading exponent $\lambda_1$, equivalent to the average fermion energy gap. The gap diverges to infinite when $t\to \pi/4$ because the subleading eigenvalues of Kraus operator all vanish. 
   (c) The blue dots show the gauge trajectory entropy. At $t=\pi/4$ it is analytically solved to be $S=2/(3(1+1/L))\ln 2$. Note that each measurement trajectory consists of $N(r+1)/2$ bits. 
   The red lines denote the entropy if the state is further evolved up to asymptotic long time, which is dominated by the leading Lyapunov exponent $S\to -2\lambda_0/N$. Note that the blue data and the red lines agree at the large $t$ regime because of a large diverging Lyapunov gap when $t\to \pi/4$, such that the leading Lyapunov exponent already dominates the trajectory entropy. 
   }
   \label{fig:Lyapunov}
\end{figure}

The Lyapunov exponent of the purification dynamics can be related to the eigen energy of the effective Hamiltonian: 
\begin{equation}
\lambda_0 = \sum_{\mathbf{su}}p_{\mathbf{su}} \left(-\min \text{eig} \left(\frac{1}{4} \mathbf{c} H_{\mathbf{su}} \mathbf{c}\right)\right) -\frac{1}{\beta}\ln B= -E_0 - \frac{1}{\beta}\ln B \ ,
\end{equation}
where $E_0$ is the {\it ground state energy}: $E_0 = \frac{1}{2}\sum_{\epsilon\leq 0} \epsilon_{n}$, averaged over $\mathbf{su}$, which converges to the average energy $E$ at asymptotic long time. The numerical results are shown in Fig.~\ref{fig:Lyapunov}a.  

The ensemble of net gauge trajectory consists of $N(r+1)/2$ bits defined over the Kekulé bonds in the 2+1D spacetime bulk. Its mean Shannon entropy can be obtained as
\begin{equation}
\begin{split}
S_{su} &= -\sum_{\{su\}}p_{\mathbf{su}} \ln p_{\mathbf{su}} = - [ \ln p_{\mathbf{s}} ]
= [\beta F_{\mathbf{s}}] +\ln B \ ,
\end{split}
\end{equation}
where $\mathbf{u}$ is absorbed by $\mathbf{s}$. Thus the average Majorana free energy has the physical meaning as the Shannon entropy of the net gauge trajectory, as shown in Fig.~\ref{fig:Lyapunov}c.

\subsection{Edwards Anderson correlations}
Consider generic observables that can be expressed as $g_{\mathbf{u}} f_{\mathbf{su}}$ where $f_{\mathbf{su}}$ can represent the Majorana correlation functions or energies that only depend on the net gauge field, while $g_{\mathbf{u}}$ represents the gauge strings that satisfy $g_{\mathbf{uu'}}=g_{\mathbf{u}}g_{\mathbf{u'}}$. 
\begin{equation}
\begin{split}
[\langle g_{\mathbf{u}} f_{\mathbf{su}} \rangle ] 
\propto& 
\sum_{\mathbf{s}} P(\mathbf{s}) \sum_{\mathbf{u}} \frac{p_{\mathbf{su}}}{P(\mathbf{s})} 
g_{\mathbf{u}} f_{\mathbf{su}}
\propto\left(\sum_{\mathbf{u}}g_{\mathbf{u}}\right)=0\ .
\end{split}
\end{equation} 
Consequently, the measurement average of any linear function of the gauge field $g_{\mathbf{u}}$ vanish exactly. 
The physical reason is that the probability depends on the net gauge field experienced by Majorana fermions, and thus the internal gauge field can always be screened by proper external gauge field trajectory. Thus upon averaging the external gauge trajectory first, different internal gauge field configurations always share equal probability. 
Specifically, this means any open Wilson line, or flux becomes strictly zero upon average. 

Then let us consider the second moment i.e. Edwards-Anderson correlation:
\begin{equation}
\begin{split}
[\langle g_{\mathbf{u}} f_{\mathbf{su}} \rangle^2 ] 
&=
\sum_{\mathbf{s}} P(\mathbf{s}) \left(\sum_{\mathbf{u}} \frac{p_{\mathbf{su}}}{P(\mathbf{s})} 
g_{\mathbf{u}} f_{\mathbf{su}}\right) \left(\sum_{\mathbf{u'}} \frac{p_{\mathbf{su'}}}{P(\mathbf{s})} 
g_{\mathbf{u'}} f_{\mathbf{su'}}\right)\\
&=
\sum_{\mathbf{s}} p_{\mathbf{s}} \sum_{\mathbf{u}} \frac{p_{\mathbf{su}}}{P(\mathbf{s})} 
 f_{\mathbf{s}}g_{\mathbf{u}}f_{\mathbf{su}}.
\end{split}
\end{equation} 
In this way one can first sample the gauge trajectory $\mathbf{s}$ according to $p_{\mathbf{s}}$, and then sample the internal {\it static} gauge field configuration using Markov chain according to their relative probability $p_{\mathbf{su}}/p_{\mathbf{s}}$. The second moment EA correlation is the correlation between two trajectories, akin to the correlation between two replicas in spin glass. The analogue specific heat for internal gauge field is the variance of the (gauge invariant) energy (defined as minus logarithm of the unnormalized Gaussian density matrix) among the internal gauge space, averaged over the measurement outcomes.

\subsection{Fermionic entanglement negativity}

\begin{figure}[h!] 
   \centering
   \includegraphics[width=\columnwidth]{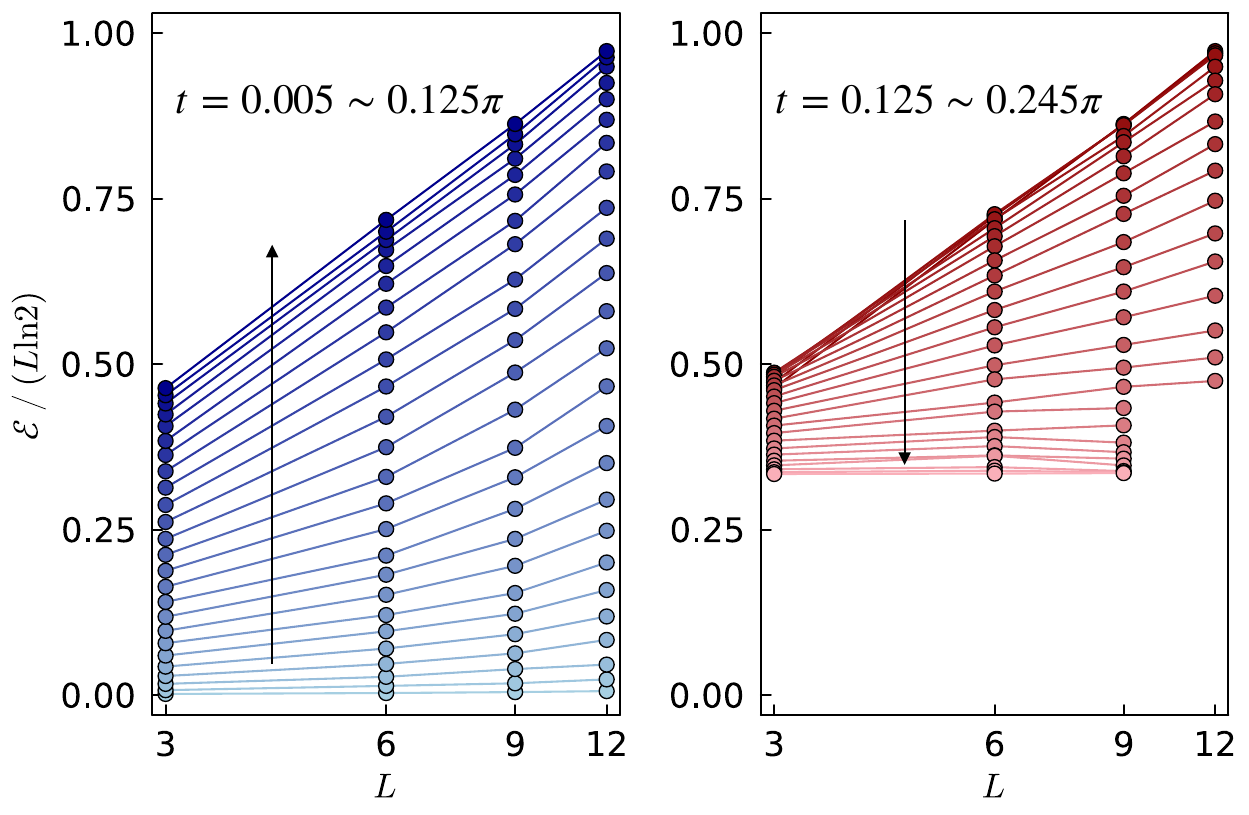} 
   \caption{{\bf Scaling of fermionic entanglement negativity across the phase diagram}.
   For the weak monitored regime $t\lesssim0.125\pi$ the entanglement grows quickly with increasing system sizes and circuit depth with an $L\ln L$ scaling. Note that a volume law scaling is precluded due to its instability against measurement in a free fermion system~\cite{Fidkowski2021howdynamicalquantum}. The seemingly faster growth than $L\ln L$ at small $t$ is attributed to a finite size effect at the early stage of power-law growth before it saturates~\cite{Zhu21quenchkitaev}. 
   For the strong measurement regime $t\gtrsim0.125\pi$, the entanglement comes to saturate to an area law. }
   \label{fig:negativityscaling}
\end{figure}

The fermionic entanglement negativity~\cite{Ryu17entneg, Eisler_2015, Ryu19entnegT, Ryu19entneg} is defined by a partial time reversal transformation over a subsystem, in analogy to the partial transpose of bosonic density matrix. For simplicity let us fix a disorder $\mathbf{su}$ but abbreviate its notation in the following. Consider a bipartition over the lattice into $A$ and its complement $\bar{A}$, then the Majorana covariant matrix can be written into blocks as follows:
\begin{equation}
\Gamma = \left(
\begin{matrix}
\Gamma_{AA} & \Gamma_{A\bar{A}} \\
\Gamma_{\bar{A}A} & \Gamma_{\bar{A}\bar{A}} \\
\end{matrix}
\right) \ .
\end{equation} 
The {\it partial time reversal} transformed fermion density matrix (denoted as $\rho^{R_A}$) is still Gaussian (in contrast to the partial transposed density matrix), whose covariant matrix is~\cite{Eisler_2015, Ryu17entneg}:
\begin{equation}
\Gamma_+ = \left(
\begin{matrix}
-\Gamma_{AA} & +i\Gamma_{A\bar{A}} \\
+i\Gamma_{\bar{A}A} & \Gamma_{\bar{A}\bar{A}} \\
\end{matrix}
\right) \ .
\end{equation} 
Since it is not hermitian: $\Gamma_+ \neq \Gamma_+^\dag \equiv \Gamma_-  $, one can define an unnormalized hermitian Gaussian density matrix 
\begin{equation}
\rho_* \equiv \rho^{R_A} \rho^{R_A\dag}  \ ,
\end{equation} 
whose norm is $\text{Tr}\left(\rho_*\right) = \sqrt{\det\frac{\mathbb{I}+\Gamma^2}{2}}$, and its covariant matrix~\cite{Calabrese10covariant}
\begin{equation}
\Gamma_* = \mathbb{I} - (\mathbb{I}-\Gamma_-)(1+\Gamma_+\Gamma_-)^{-1}(\mathbb{I} - \Gamma_+) \ .
\end{equation} 
Denoting the eigenvalues of $\text{eig}(\Gamma_*)=\xi_n$ and $\text{eig}(\Gamma)=\zeta_n$, the fermionic entanglement negativity follows as~\cite{Ryu17entneg}
\begin{equation}
\begin{split}
\mathcal{E} &\equiv \ln \text{Tr} \sqrt{\rho^{R_A\dag}\rho^{R_A} } \\
&= \ln\frac{\text{Tr}\sqrt{\rho_*}}{\sqrt{\text{Tr}\rho_*}} +\frac{1}{2}\ln\text{Tr}\left(\rho_*\right)\\
&=
\sum_{n=1}^{N/2} \ln \left(\sqrt{\frac{1+\xi_n}{2}}+\sqrt{\frac{1-\xi_n}{2}}\right)
+
\frac{1}{2}\ln\frac{1+\zeta_n^2}{2} \ .
\end{split}
\end{equation}
The negativity scaling for each $t$ is shown in Fig.~\ref{fig:negativityscaling}. 

\subsection{Two step nested Monte Carlo sampling}

\begin{figure}[h!] 
   \centering
   \includegraphics[width=.8\columnwidth]{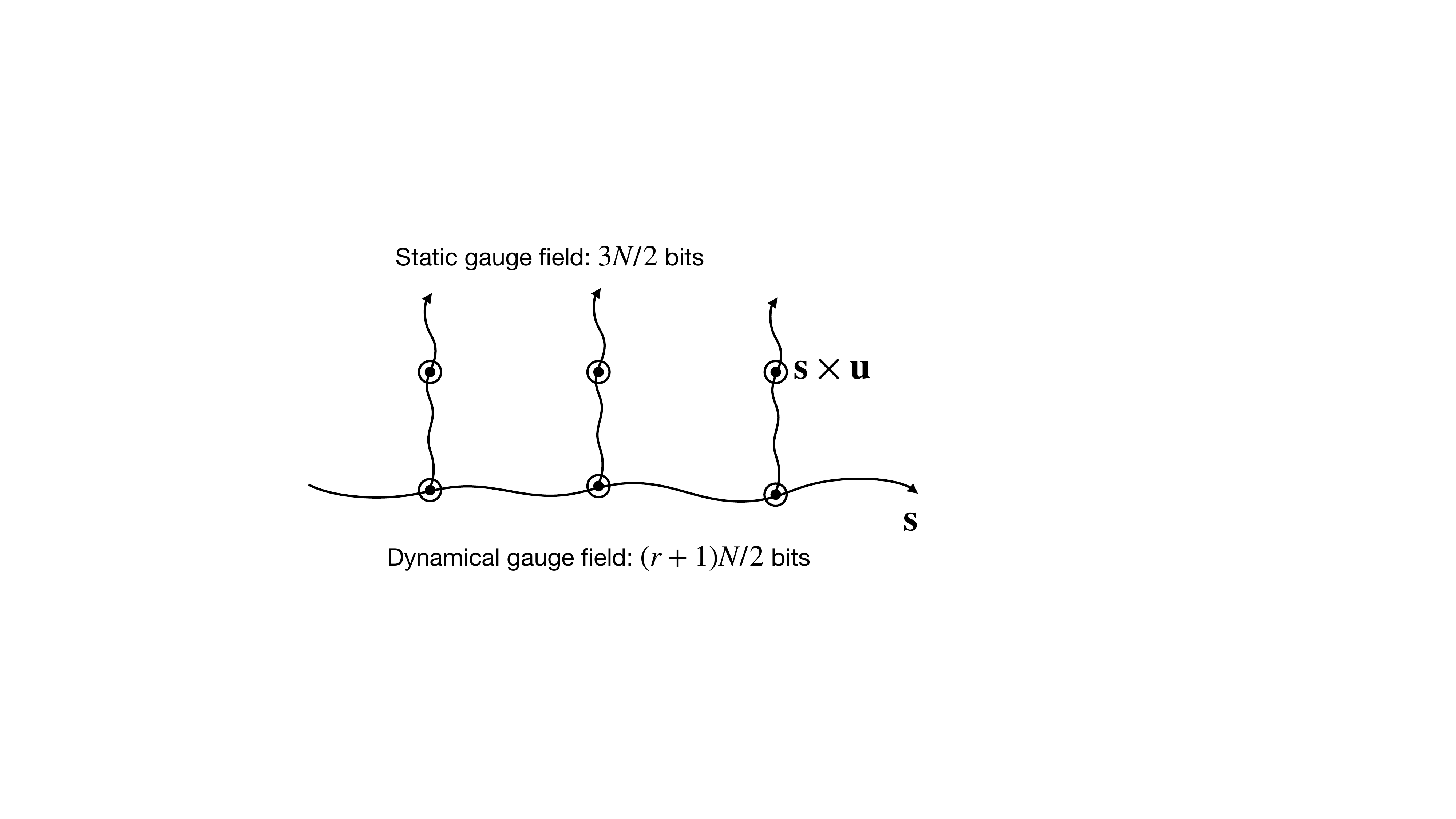} 
   \caption{Schematic of the comb-shaped Markov chains with branchings. Each point is a disordered Majorana state coupled to a spatiotemporal gauge configuration $\mathbf{su}$, whose probability is obtained by tracing out fermion. }
   \label{fig:markovchain}
\end{figure}

The Monte Carlo sampling is schematically shown in Fig.~\ref{fig:markovchain}. Horizontal chain is for the dynamical gauge trajectory $\mathbf{s}$ in a $L\times L\times (r+1)$ dimensional spacetime grid, while vertical chain is for static gauge field $\mathbf{u}$ in a $L\times L\times 3$ grid. We sample 2000 sweeps along the horizontal chain for $\mathbf{s}$, discard the first 500 sweeps for equilibrium ensemble, and branch out to grow the vertical chain in every 100 sweeps interval, and sample 1000 sweeps for $\mathbf{u}$, according to relative probability $p_{\mathbf{su}}$. We take vertical branches separated by a relatively large distance comparable or greater than the auto-correlation time. By using binning analysis to integrate out each vertical chain for an averaged physical observable $\langle \cdots\rangle \equiv \sum_{\mathbf{u}}\frac{p_{\mathbf{su}}}{P(\mathbf{s})}(\cdots)$, a Markov comb reduces to a Markov chain. Then we perform binning analysis to integrate out the reduced Markov chain $[\langle\cdots\rangle] \equiv \sum_{\mathbf{s}} p_{\mathbf{s}}\langle \cdots\rangle $ for final observables: $[\langle\cdots\rangle] \equiv \sum_{\mathbf{s}} P(\mathbf{s})\sum_{\mathbf{u}}\frac{p_{\mathbf{su}}}{P(\mathbf{s})}(\cdots)$. 

\section{Supplemental numerical data}

\subsection{Entanglement negativity of Kitaev Hamiltonian at finite temperature}
\begin{figure}[h!] 
   \centering
   \includegraphics[width=\columnwidth]{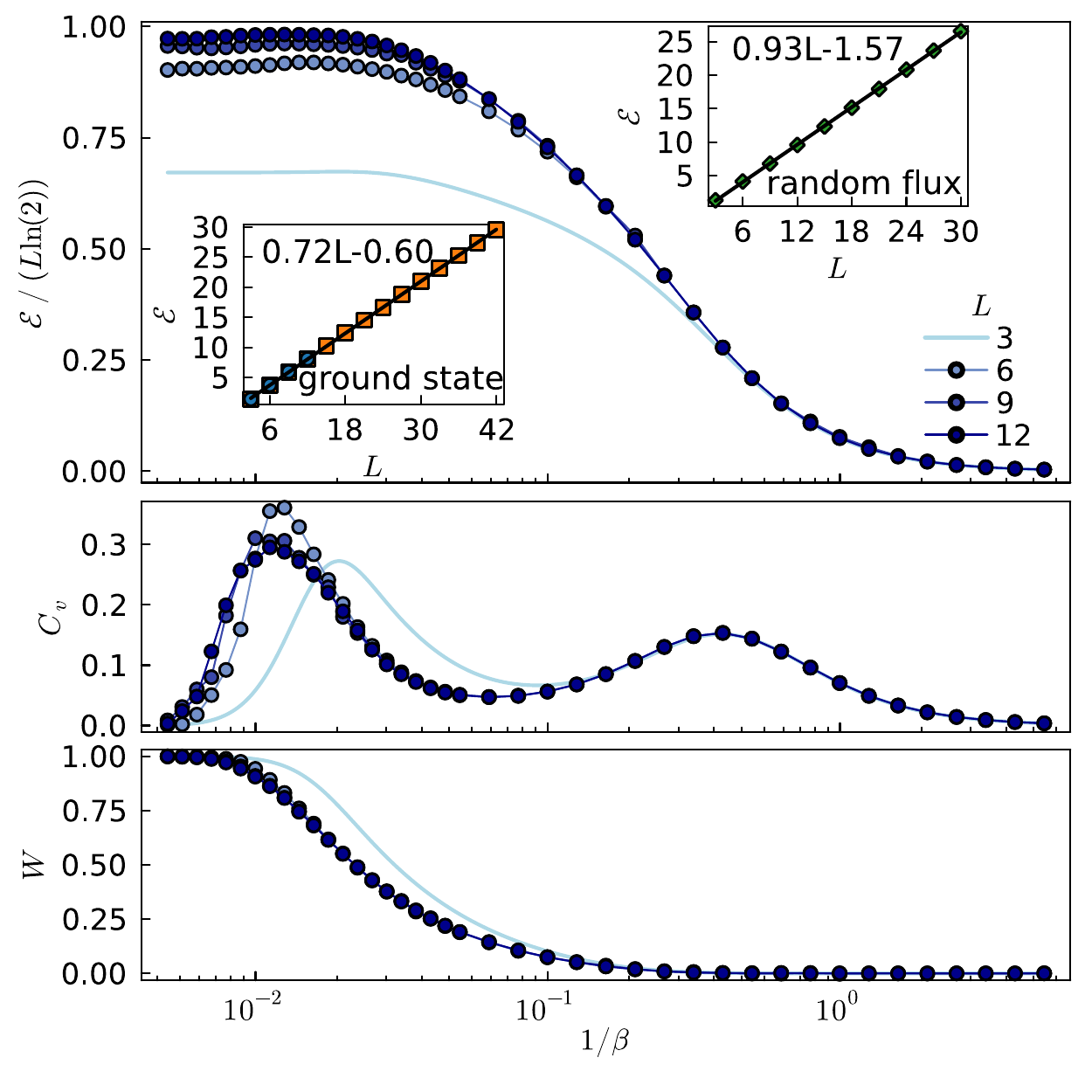} 
   \caption{{\bf Kitaev model at finite temperature $1/\beta$}. (a) Fermionic entanglement negativity, for half cut two-cylinder bipartition, which converges to an area law even down to zero temperature. In the left inset, we plot the scaling at $1/\beta=0.01$ (blue circles), in comparison with the exact ground state $1/\beta=0$ i.e. fermionic ground state under zero flux, denoted by orange squares, that obeys the area law up to large system sizes. In the right inset, we show the negativity of fermionic ground state under random flux, which also obeys area law with a different coefficient. 
   (b) Energy variance (specific heat capacity). (c) Flux. 
   The data of $L=3$ is computed by exact summation of the local fluxes. The fermion is put in anti-periodic boundary condition that has lowest energy at finite size. }
   \label{fig:kitkek}
\end{figure}

For comparison with the quantum circuit, we compute the finite temperature fermionic entanglement negativity of the following Hamiltonian 
\begin{equation}
H = \sum_{\langle ij\rangle \in R} Z_i Z_j + \sum_{\langle ij\rangle \in G} Y_i Y_j + \sum_{\langle ij\rangle \in B} X_i X_j \ ,
\label{eq:KitaevHam}
\end{equation}
which is equivalent to the conventional Kitaev Hamiltonian~\cite{Kitaev2006, Nasu15kitaevfiniteT} up to a local basis rotation~\cite{Schmidt10honeycomb, Hastings22honeycomb}, see Fig.~\ref{fig:kitkek}. 

\subsection{Random disorder compared with Born disorder}
\begin{figure}[h!] 
   \centering
   \includegraphics[width=\columnwidth]{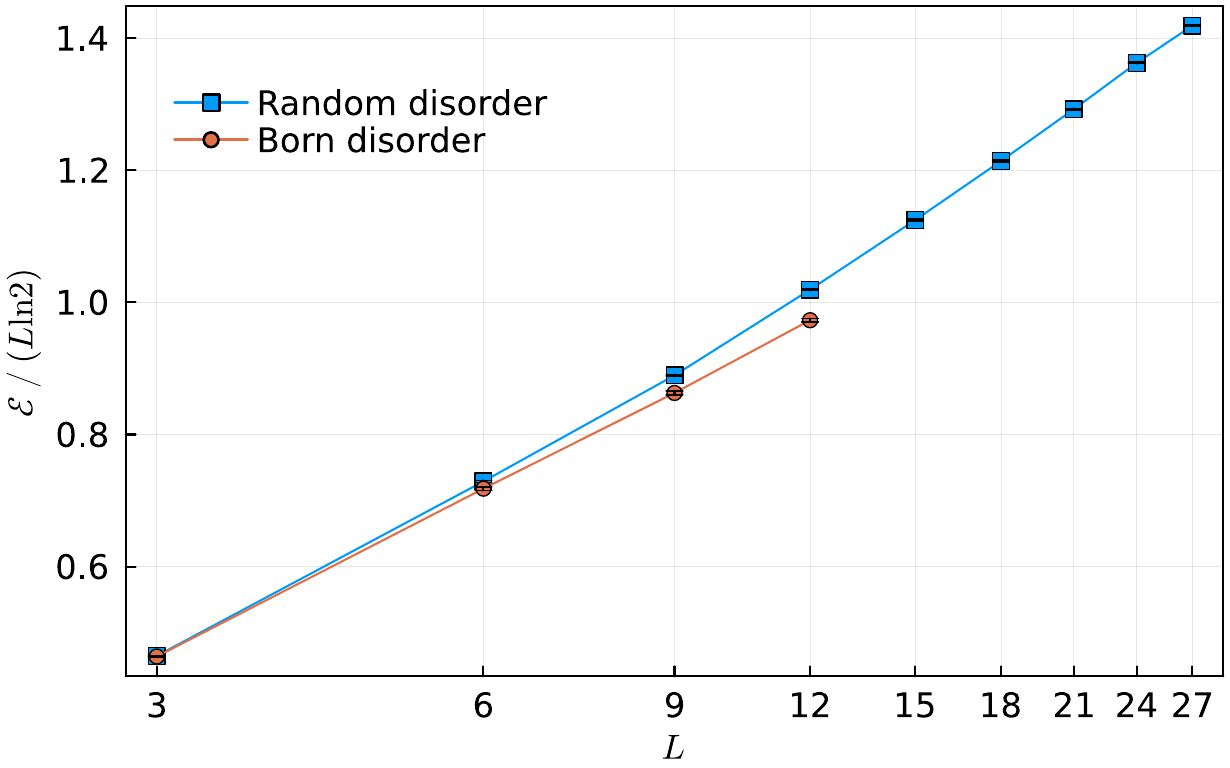} 
   \caption{{\bf Random bond disorder in spacetime versus Born probability disorder}, at $t=0.125\pi$. The typical trajectories in both cases exhibit $L\ln L$ scaling entanglement, which goes beyond the area law in a random flux Hamiltonian ground state in Fig.~\ref{fig:kitkek}. Note that the two disorder scenarios diverge when increasing system size and circuit depth, and the Born trajectory appears less entangled than the random typical trajectory.}
   \label{fig:randvsBorn}
\end{figure}

Here we compare the fermionic negativity of the fermion state in the typical random disordered $\mathbf{su}$ trajectory, compared with that in the typical Born disordered trajectory (with probability $p_{\mathbf{su}}$) we discussed in the main text, see Fig.~\ref{fig:randvsBorn}. 

\clearpage

\end{document}